\documentclass[11pt,fleqn]{article}

\usepackage{bm}
\usepackage{amsmath}
\usepackage{mathtools}
\usepackage{graphicx}
\usepackage{cite}
\usepackage[margin=1.25in]{geometry}
\usepackage{booktabs}
\usepackage{caption}
\usepackage{subcaption}
\usepackage{nicefrac}
\usepackage{placeins}
\usepackage{afterpage}
\usepackage{float}
\usepackage[arrowdel]{physics}

\usepackage{color}

\usepackage{color}
\definecolor{olivedrab}{rgb}{0.42,0.56,0.14}
\definecolor{oxfordblue}{rgb}{0.0, 0.13, 0.28}
\definecolor{arabianblue}{rgb}{0.1, 0.3, 0.61}
\definecolor{oldmauve}{rgb}{0.4, 0.19, 0.28}

\newcommand{\tensr}[1]{\bm{\mathsf{#1}}}

\newcommand{\sss}{\scriptscriptstyle}
\newcommand{\sbs}[1]{_{\sss#1}}
\newcommand{\sps}[1]{^{\sss#1}}

\newcommand{\K}[1]{\kappa\sbs{#1}}
\newcommand{\Keq}[1]{\kappa \sbs{#1}\sps{eq}}

\newcommand{\wphi}[1]{\omega\sbs{#1}\sps{\phi}}

\graphicspath{{figures/}} 

\newcommand{\um}{\scalebox{0.75}[1.0]{\( - \)}}

\newcommand{\PP}{\tensr{P}}
\newcommand{\F}{\tensr{F}}

\makeatletter
\newcommand{\mathleft}{\@fleqntrue\@mathmargin0pt}
\newcommand{\mathcenter}{\@fleqnfalse}
\makeatother

\graphicspath{{figures/}} 

\title{\vspace{-2.0cm} Investigation of coalescence and pinch-off processes of a self-rewetting drop interacting with a liquid layer}

\author{{Bashir Elbousefi, William Schupbach, Kannan N. Premnath, Samuel W.J. Welch}\\Department of Mechanical Engineering\\ University of Colorado Denver\\ 1200 Larimer Street, Denver, CO 80204, U.S.A.}

\begin{document}

\maketitle

\begin{abstract}
Self-rewetting fluids (SRFs), such as long-chain alcohol solutions, are a special class of liquids with surface tension that is anomalously depends quadratically on temperature, resulting in thermocapillary flows that differ significantly from those in normal fluids (NFs). Recent interest in SRFs are mainly due to their role in enhancing fluid dynamical and thermal transport in various microgravity applications and microfluidics, while much of their fundamental processes remain unexplored. This study focuses on simulating and investigating the behavior of SRF drops interacting with a self-rewetting liquid layer under nonuniform heating conditions. In this regard, we employ a robust central moment-based lattice Boltzmann method (LBM) with a phase-field model, which incorporates three distribution functions: one for two-fluid motion with high-density ratios including the interfacial Marangoni stresses, another for interface capturing based on the conservative Allen-Cahn equation, and a third for the energy equation expressed in an axisymmetric formulation to capture three-dimensional effects efficiently. We investigate the coalescence and pinch-off processes in SRFs and compare them to those in NFs. Our simulations reveal that SRFs undergo pinch-off earlier than NFs. In SRFs, the fluid moves toward the hotter region around interfaces, which is opposite to the flow in NFs. We also observe that increasing the Ohnesorge number $\mbox{Oh}$ suppresses the pinch-off process, highlighting the role of viscous forces relative to surface tension, which is modulated by gravity effects or the Bond number $\mbox{Bo}$. Furthermore, we explore how varying the dimensionless linear and quadratic sensitivity coefficients of surface tension on temperature, $M_1$ and $M_2$, respectively, and the dimensionless heat flux $Q$ influence the coalescence/pinch-off behavior. Interestingly, in SRFs increasing $M_2$ or $Q$ reduces the time required to pinch-off and widens the region in the $\mbox{Oh}$-$\mbox{Bo}$ regime map where pinch-off occurs, when compared to the unheated cases; by contrast, in NFs, increasing $M_1$ or $Q$ extends the residence time prior to pinch-off and widens the region in the $\mbox{Oh}$-$\mbox{Bo}$ map where coalescence occurs. These differences are shown to be due to the variations in the thermocapillary forces on the interface. Overall, we find that under nonuniform heating, the SRFs enhance the pinch-off process, resulting in shorter pinch-off times compared to NFs across a wider range of conditions.

\end{abstract}

\section{Introduction}
The coalescence of a liquid drop on a liquid pool has long been of interest due to its fundamental relevance to a variety of natural and industrial systems and its importance in many physical processes, including raindrop dynamics and droplet formation in clouds~\cite{berry1974analysis}, droplet emulsion~\cite{cockbain1953stability}, microfluidics applications~\cite{velev2003chip, stone2004engineering}, separation of oil phase from water phase in emulsion~\cite{hirato1991demulsification}, and precipitation (production of ocean mists containing salt particles)~\cite{raes2000formation}, spray atomization in combustion engines~\cite{hirato1991demulsification}, and spray cooling heat transfer~\cite{kim2007spray}.

The main mechanisms for coalescence of the drop depend on the strength of surface tension, gravity, and viscous forces where the drop may undergo complete or partial coalescence (see e.g.~\cite{charles1960mechanism,blanchette2006partial}). In the coalescence process of the liquid drop on a liquid pool, a neck develops at the contact point of the drop and liquid pool. The neck expands fast due to capillary pressure. The expanding neck generates capillary waves that propagate along the drop interface. Simultaneously, the drop liquid drains into the pool due to the excess capillary pressure inside the drop. The draining drop converts to a cylindrical protrusion as the capillary waves converge at the apex of the drop. After reaching a maximum point, the apex of the cylindrical column retracts vertically, while its neck retracts horizontally. Depending on the opposition among those horizontal and vertical retraction rates, the most important topological adjustments of the drop are observed above the interface: partial coalescence~\cite{charles1960mechanism, thoroddsen2000coalescence, mohamed2003drop,mohamed2004drop,aryafar2006drop,honey2006astonishing,chen2006partial, blanchette2006partial, blanchette2009dynamics, geri2017thermal,sun2018marangoni, deka2019coalescence,alhareth2020partial}, in which the liquid drop coalesces partially with the pool leaving behind secondary droplets, and complete coalescence~\cite{thomson1886v,chapman1967formation,rodriguez1985some,rodriguez1988penetration,peck1995vortex,shankar1995vortex,durst1996penetration,saylor2003effect,lee2015origin,behera2019generation,behera2021viscous}, in which the liquid drop merges completely with the pool without producing any daughter drop.

Numerous multiphase and thermal transport mechanisms significantly rely on surface tension forces generated at the fluid-fluid interface~\cite{de2004capillarity}. Changes in the local interfacial temperature or the addition of surface-active substances (such as surfactants) can lead to local variations in them. The so-called Marangoni stresses, which are produced by the surface tension gradients; see~\cite{scriven1960marangoni}, are caused by the fluids' viscous effects and cause convective motions close to the interfaces~\cite{probstein2005physicochemical}. They are known as thermocapillary convection if they form due to local temperature variations. These differences in temperature between two fluid mediums can prevent the coalescence and promote a rebound effect aided by an intervening layer of air. The thermally-induced capillarity stemming from the temperature gradient can lead to the separation of droplets in various circumstances, such as when different liquids come into contact. The temperature-dependent surface tension of the liquids can create Marangoni tangential stresses at the fluid interfaces. These Marangoni stresses can push air into the layer between the liquids, thereby bolstering the stability of the air layer and delaying the merging process~\cite{geri2017thermal}. The temperature difference between the liquids has a direct influence on the occurrence of this phenomenon. Furthermore, the temperature difference at which coalescence is delayed is associated with the viscosity of the liquids. As a result, employing the temperature gradient technique proves effective in enhancing the rebounding action between a droplet and a liquid reservoir.

The surface tension of most fluids has the characteristic of decreasing approximately linearly with temperature. On the other hand, some fluids, such as aqueous solutions of long-chain (i.e., "fatty") alcohols (such as n-butanol, pentanol, and heptanol to name new), some liquid metallic alloys, and nematic liquid crystals, show anomalous nonlinear parabolic dependence of surface tension on temperature with a range involving its positive gradient and are known as self-rewetting fluids (SRFs) (see e.g.~\cite{vochten1973study,petre1984experimental,villers1988temperature}). As a result, the self-rewetting fluids can provide a strong inflow of liquids towards high-temperature areas, such as towards nucleating sites during boiling, preventing the formation of dry patches at such hot locations. Over the past two decades, research into these unique classes of fluids to improve transport in various thermal management applications has been driven by these and other distinctive properties resulting from the thermocapillarity associated with SRFs. Both terrestrial and microgravity situations have been suggested as working fluids for different technical purposes~\cite{abe2004microgravity,abe2007terrestrial}. It has been demonstrated that using SRFs increases the effectiveness of heat transfer in heat pipes, flow boiling and evaporation in microchannels, pool boiling processes, and two-phase heat transfer devices using self-rewetting gold nanofluids~\cite{zaaroura2021thermal,savino2009surface,savino2013some,hu2014heat,wu2017study, cecere2018experimental,zhu2020thermal}. We have recently provided insights into selected configurations involving SRFs~\cite{elbousefi2024lattice} such as in superimposed fluid layers by developing new analytical solution and a numerical approach~\cite{elbousefi2023thermocapillary} and in bubble migration laden with surfactants via simulations~\cite{elbousefi2024investigation}.

Various studies on the phenomena of drop coalescence on a heated liquid layer have applications in various real-world scenarios, such as in oil industry, bioprinting, manufacturing nanoparticles, and microfluidic technologies, are discussed in the literature~\cite{abdelaziz2013green,schwenzer2014leidenfrost,geri2017thermal,mrinal2017self,mogilevskiy2020levitation, kirar2020coalescence,poureslami2023evaporative}. In the scenario of normal fluids (NFs) where the surface tension decreases linearly with temperature, Marangoni stress generated interfacial fluid dynamics has been investigated in some numerical and experimental studies to have a considerable impact on droplet coalescence in a pool~\cite{blanchette2009influence,blanchette2010simulation,martin2015simulations,kim2015spontaneous,sun2018marangoni,jia2020marangoni}.
The Marangoni flow on the droplet impingement process can become more complex if the surface tension exhibits a quadratic relationship with respect to temperature as in SRFs which have been used in various applications recently as discussed above, as opposed to normal fluids where the surface tension gradient is always negative. Studying the flow physics and thermal transport behind such possibilities have not yet been fully explored. One of the main goals of the current work is to numerically simulate and analyze the coalescence or pinch-off processes of a self-rewetting drop interacting with a self-rewetting liquid layer that is subject to nonuniform heating represented using a Gaussian profile for the surface heat flux variation or an imposed nonuniform temperature on the heater surface and compare them with those based on normal fluids. The base heating is introduced so as to naturally generate and sustain a temperature gradient and consequently the Marangoni effects in the vicinity of the fluid interfaces at longer times; as a result, this represents a more realistic representation and a general situation of the thermocapillary flows and attendant coalescence and pinch-off phenomena in SRFs induced due to the presence of a heater.

In this regard, we will utilize a numerical simulation method based on a robust central moment lattice Boltzmann (LB) approach. This method is based on a phase field model derived from the conservative Allen–Cahn equation, which is an extension and enhancement of our prior work~\cite{hajabdollahi2021central,elbousefi2023thermocapillary,elbousefi2024investigation}. This approach involves calculating the evolution of three distinct distribution functions via the standard collide-and-stream approach in the LB framework: one for the flow field, another for the temperature field, and the third for capturing interfaces through an order parameter and with an attendant surface tension equation of state for SRFs. To efficiently account for three-dimensional effects, we have implemented this phase field-based central moment LB approach for simulating interfacial flows in an axisymmetric formulation.

The structure of this paper is as follows. The problem setup of the coalescence and pinch-off of a self-rewetting drop onto a liquid layer will be covered in the next section (Section~\ref{DOLI HF Section.1}), where the governing equations, the non-dimensional groups and surface tension interface equation of state used in this work are presented in Subsection~\ref{DOLI HF Subsection.1.2}. Section~\ref{Sec.3} contains the computational model equations for the LB schemes for multiphase flows utilizing a phase field model. In this regard, the key ideas of the LBM implementations and the discretized central moment LB algorithms for modeling multiphase flows with thermal transport in SRFs are discussed in Sec.~\ref{Sec.LBschemes}; their extension to simulations in axisymmetric coordinates using a quasi two-dimensional formulation with source terms are briefly outlined in Sec.~\ref{DOLI HF Axisymmetric Model}. A numerical validation of our computational approach using a previously established benchmark problem of the drop interaction with a liquid layer involving a normal fluid for the unheated case is discussed in Sec.~\ref{DOLI Benchmark}. Section~\ref{DOLI HF Simulation Results and Discussion} presents the simulation results and discussion of the effect of various characteristic parameters on the thermocapillary modulated coalescence and pinch-off of a self-rewetting drop into a SRF liquid layer subjected to an imposed heat flux and its comparison with the corresponding normal fluid cases. This is followed by Sec.~\ref{Temp BC}, which provides additional results and discussion for the case where the heating and thermocapillary effects are achieved by an imposed nonuniform temperature distribution. Finally, the main conclusions of this work are summarized in Sec.~\ref{Sec.9 HF}.

\section{Problem statement, governing equations, and dimensionless parameters} \label{DOLI HF Section.1}
\begin{figure}[H]
\centering
\includegraphics[trim = 0 0 0 0, clip, width =90mm]{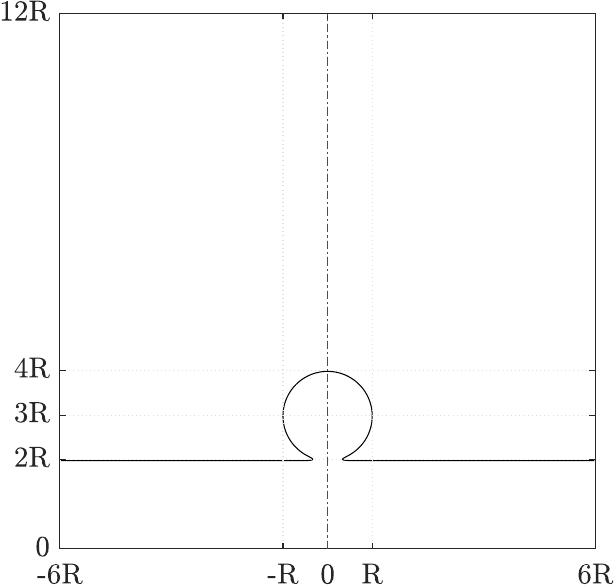}
\caption{Schematic of the initial condition for a drop of radius R on a liquid interface for the non-heated case. Using an axisymmetric model, the axis of symmetry is the vertical centerline shown in the figure, and the entire domain is shown here for clarity.}
\label{DOLI_Initial_Condition}
\end{figure}

\begin{figure}[H]
\centering
\includegraphics[trim = 0 0 0 0, clip, width =90mm]{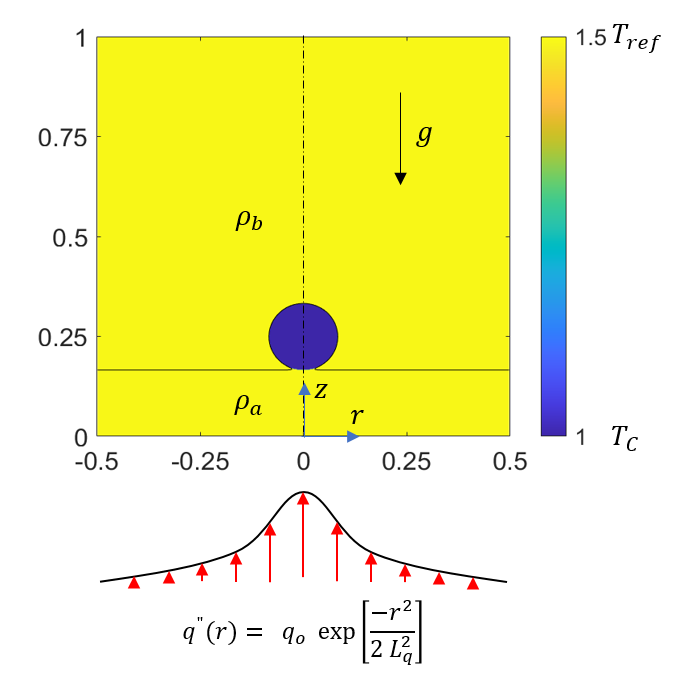}
\caption{Schematic of the initial condition for a drop of radius R on a liquid interface for the heated case. Using an axisymmetric model, the axis of symmetry is the vertical centerline shown in the figure, and the entire domain is shown here for clarity. The heat flux $q^{\prime\prime}(r)$ imposed at its bottom side is given by a Gaussian function with a magnitude $q_o$ and a horizontal spread length scale $L_q$, which in turn induces nonuniform heating and thermocapillary effects on the interface.}
\label{HFBC_heated_MODEL}
\end{figure}

In this numerical simulation study, our goal is to perform axisymmetric simulations of the coalescence/pinch-off of a drop interacting with a SRF liquid layer under nonuniform heating conditions, similar to that found in \cite{blanchette2006partial}, which analyzed this case with a NF under unheated conditions. A drop of radius, $R$, is placed above a flat liquid interface such that a neck forms that initiates the coalescence process at the initial condition. Then, under the gravity force, the drop sinks into and coalesces with the liquid pool; however, when the surface tension is large compared to the viscous stress, a secondary drop will form. On the other hand, if the surface tension is small compared to the viscous stress, the secondary droplet will not form, and the initial drop will become completely absorbed by the liquid pool.
We set this problem up such that the boundaries are far away from the drop, similar to~\cite{constante2021role} so that they do not influence the physical behavior of the coalescence process. Here, we use three solid walls (top, bottom, and side) and a wet no-gradient boundary representing the axis. The problem configurations are illustrated in Figs.~\ref{DOLI_Initial_Condition} and~\ref{HFBC_heated_MODEL} for the non-heated and heated cases, respectively, where fluid $a$ denotes both the drop and the liquid interface, whereas the surrounding medium is indicated by fluid $b$. The dynamic viscosities and the densities of fluid $a$ are denoted by $\mu_a$ and $\rho_a$, respectively, while those for fluid $b$ are represented by $\mu_b$ and $\rho_b$, respectively.
The size of the computational domain resolving along the radial and axial directions in the simulation is $(12R\times12R)$ with a liquid layer filled in its bottom side such that initially flat interface is located at a distance of $2R$ from the axis below, and the center of the drop is initially located at a distance of $3R$ from the axis below.

For the heated case, we set up the heating conditions such that the drop is initially at a cooler temperature $T_C$ in an ambient medium maintained at a temperature $T_{ref}$ and the liquid layer is subject to a nonuniform heating in its bottom side at $z=0$ represented using a Gaussian profile for the heat flux variation (see Fig.~\ref{HFBC_heated_MODEL}). In this regard, the heat flux is characterized by a magnitude $q_o$ at the center $r=0$ and its horizontal spread length scale in its normal distribution is taken as $L_q$ (see Eq.~(\ref{eq:Gaussianheatfluxprofile}) for details).

\subsection{Governing equations, surface tension interface equation of state, and non-dimensional groups} \label{DOLI HF Subsection.1.2}
The thermocapillary flow generated by nonuniform interfacial heating in the SRFs obeys the equations of mass and momentum (i.e., Navier–Stokes equations (NSE)) and energy transport, which can be respectively written as follows:
\mathleft
\begin{subequations}
\begin{equation}
\qquad \bm{\nabla} \cdot {\bm{u}}=0,
\end{equation}
\begin{equation}
\qquad \rho \left( \frac{\partial \bm{u}}{\partial t} + \bm{u}\cdot \bm{\nabla}\bm{u} \right) = - \bm{\nabla} p + \bm{\nabla} \cdot \left[ \mu (\bm{\nabla} \bm{u} + \bm{\nabla} \bm{u}^{\dagger})\right] + \bm{F}_{ext},
\end{equation}
\begin{equation} \label{energy eqn}
\qquad \frac{\partial {T}}{\partial t} + \bm{u} \cdot \bm{\nabla}T = \bm{\nabla} \cdot \left(\alpha \bm{\nabla}T \right),
\end{equation}
\end{subequations}
where $\rho$, $\mu$, and $\alpha$ stand for the fluid density, dynamic viscosity, and thermal diffusivity of the fluid, respectively. The value of $\alpha$ is determined by the ratio of thermal conductivity $k$ to specific heat $c_p$, as $\alpha=k/(\rho c_p)$. In the above, $\bm{u}$, $p$, and $T$ represent the velocity, pressure, and temperature fields of the fluids, respectively, and the superscript symbol $\dagger$ signifies taking the transpose of the dyadic velocity gradient $\bm{\nabla} \bm{u}$. In addition, $\bm{F}_{ext}$ is an external force such as gravity.

\begin{figure}[H]
\centering
\includegraphics[trim = 0 0 0 0, clip, width =80mm]{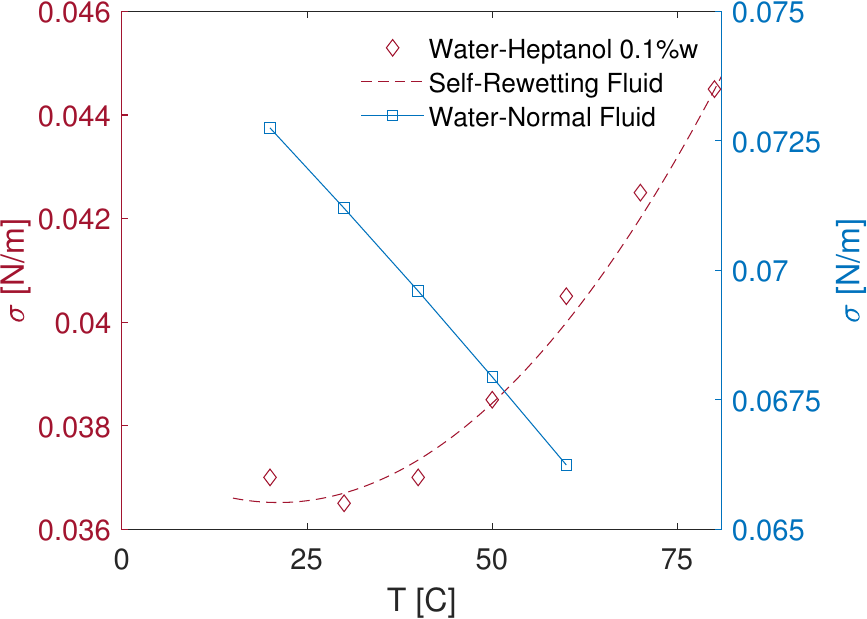}
\caption{Illustration of the variation of the surface tension with temperature for a normal fluid (NF) and a self-rewetting fluid (SRF). For the SRF, the surface tension exhibits a quadratic polynomial dependence on temperature, derived through a curve fitting of the experimental data reported by Savino et al.~\cite {savino2009surface}, with a positive gradient for the temperature range considered.}
\label{fig:SRF_vs_NF_Illustration}
\end{figure}

In order to establish an equation governing the surface tension at the interface, we need to relate it to the variations in the local temperature, denoted as $T$, that effectively models the behavior in self-rewetting fluids (SRFs). In this regard, we utilize \emph{a nonlinear (quadratic) relationship between surface tension and temperature}, as given in the following equation:
\mathleft
\begin{equation}\label{ST_SRF}
\qquad \sigma (T) = \sigma_o + \sigma_T (T-T_{ref})+ \sigma_{TT} (T-T_{ref})^2,
\end{equation}
where $\sigma_o$ denotes the local value of the surface tension at a reference temperature $T_{ref}$, $\sigma_{T}=\frac{d\sigma}{dT}\big\vert_{T_{ref}}$ and $\sigma_{TT}=\frac{1}{2}\frac{d^2\sigma}{dT^2}\big\vert_{T_{ref}}$ are the surface tension linear and quadratic sensitivity coefficients, respectively, expressing the sensitivity of the surface tension on temperature. It is important to note that in the case of a SRF, $\sigma_{TT}\neq 0$. On the other hand, for a NF, $\sigma_{TT} = 0$ where only $\sigma_T$ is non-zero. In a general context, parameters such as $\sigma$, $T_{ref}$, $\sigma_T$, and $\sigma_{TT}$ are distinctive attributes specific to a chosen SRF. Equation~(\ref{ST_SRF}) represents a best-fit representation of the variation of the interfacial equation of state of SRFs observed in various experiments (see e.g.~\cite{vochten1973study,petre1984experimental,villers1988temperature}) and widely used in their modeling and simulations (see e.g.,~\cite{tripathi2015non,elbousefi2023thermocapillary}). See Fig.~\ref{fig:SRF_vs_NF_Illustration} for an illustration of the functional dependence of the surface tension on the temperature for these fluids. Clearly, it can be seen that the NF has a negative slope while the SRF has a positive slope in the temperature range considered; the latter is crucial in realizing the self-rewetting behavior in thermocapillary phenomena.

Furthermore, when the above governing equations are non-dimensionalized using a reference velocity scale $U_{o}$ (see below) and a length scale $R$ corresponding to the radius of the drop, the following principal dimensionless numbers that characterize the coalescence of a drop onto a SRF liquid interface can be obtained: the Bond number $\mbox{Bo}$ which is used to represent the magnitude of the gravitational force relative to the surface tension force and can be written as
\begin{equation}
\qquad \mbox{Bo} = \frac{g \rho_a R^2}{\sigma_o}, \nonumber
\end{equation}
and the Ohnesorge number $\mbox{Oh}$, which is used to characterize the relative magnitude between the viscous forces and the surface tension force on the interface and is represented by \begin{equation}
\qquad \mbox{Oh} =\frac{\mu_a}{\sqrt{\rho_a \sigma_o R}}, \nonumber
\end{equation}
where, as noted above, the subscript $a$ refers to the properties of the liquid phase representing the drop and the pool.
We also define a characteristic velocity $U_o$ relevant for capillary-driven motions in terms of the surface tension, density, and size of the drop as
\begin{equation}
\qquad U_{o} = \sqrt{\frac{\sigma_o}{\rho_a R}},\nonumber
\end{equation}
which can then be used to reinterpret the Ohnesorge number as the inverse of the Reynolds number given by $\mbox{Re}=\rho_aU_oR/\mu_a$. As a consequence, we can introduce a characteristic time scale as $T_o = {R}/{U_o}$. 

In addition, the density and dynamic viscosity ratios that influence the fluid dynamical process under consideration can be written as
\mathleft
\begin{equation*}
\qquad \tilde{\rho} = \frac{\rho_a}{\rho_b}, \quad \quad \tilde{\mu} = \frac{\mu_a}{\mu_b}. \label{three}
\end{equation*}
Of special importance in characterizing the SRF effect are the dimensionless surface tension linear and quadratic sensitivity coefficients for the equation of state for the surface tension given by
\begin{equation}
\qquad M_1 =\left(\frac{\Delta T}{\sigma_o}\right)\sigma_T,\quad M_2 =\left(\frac{\Delta T^2}{\sigma_o}\right)\sigma_{TT}, \label{three-prime}
\end{equation}
where $\Delta T$ is a reference scale for the temperature difference. Note that since $\sigma_T$ and $\sigma_{TT}$ as defined in the paragraph following Eq.~(\ref{ST_SRF}) are the first and second derivatives (or sensitivity), respectively, of the surface tension with respect to the temperature, the parameters $M_1$ and $M_2$, as defined above, are dimensionless.

As mentioned earlier, the liquid layer is subject to a nonuniform heating parameterized using a Gaussian profile for the surface heat flux variation $q^{\prime\prime}(r)$ imposed on its bottom side at $z=0$, which can be represented as
\mathleft
\begin{equation}\label{eq:Gaussianheatfluxprofile}
\qquad q^{\prime\prime}(r) = q_o \exp \left[ -{\frac{r^2}{2 {L^2_q}}} \right],
\end{equation}
where $q_o$ is the maximum heat flux value at the center ($r=0$) and $L_q$ is the characteristic width of the profile. Then, a dimensionless heat flux can be defined as
\mathleft
\begin{equation}
\qquad Q = \frac{q_o L}{k_a \Delta T},
\end{equation}
where $L$ is the length of the computational domain, $k_a$ is the thermal conductivity for the liquid phase, and $\Delta T$ is a scale for temperature difference reference scale.

\section{Computational modeling} \label{Sec.3}
We will now discuss a modeling formulation suitable for developing a numerical approach based on the LBM for the simulation of interfacial dynamics involving thermocapillary flow in SRFs presented in the next section. The phase-field lattice Boltzmann approach based on the conservative Allen-Cahn equation (ACE)~\cite{chiu2011conservative} is considered in this study to capture interfacial dynamics while maintaining the segregation of two immiscible fluids. The binary fluids are distinguished by an order parameter or the phase field variable $\phi$. The fluid $a$ is identified by $\phi = \phi_a$, while fluid $b$ by $\phi = \phi_b$. The interface-tracking equation based on the conservative ACE in terms of the phase field variable is given as
\mathleft
\begin{equation}\label{eqn1}
\qquad \frac{\partial \phi}{\partial t} + \bm{\nabla} \cdot (\phi \bm{u}) = \bm{\nabla} \cdot [M_\phi(\bm{\nabla} \phi - \theta \bm{n})],
\end{equation}
where $\bm{u}$ is the fluid velocity, $M_{\phi}$ is the mobility, and $\bm{n}$ is the unit normal vector, which can be calculated using the order parameter $\phi$ as $\bm{n} = \bm{\nabla}{\phi}/|\bm{\nabla}{\phi}|$. Here, the parameter $\theta$ can be expressed as
$\theta = -4\left(\phi - \phi_{A} \right) \left(\phi - \phi_{B} \right)/[W \left(\phi_{A} - \phi_{B} \right)]$, where $W$ is the width of the interface. Essentially, the term $M_\phi\theta \bm{n}$ in Eq.~(\ref{eqn1}) serves as the interface sharpening term counteracting the diffusive flux $-M_\phi\bm{\nabla} \phi$ following the advection of $\phi$ by the fluid velocity. At equilibrium, the conservative ACE reduces the order parameter to a hyperbolic tangent profile across the diffuse interface, which is given by $\phi\left( \zeta \right)= \frac {1}{2}\left(\phi_{A} + \phi_{B} \right)+ \frac {1}{2}\left(\phi_{A} - \phi_{B} \right)\tanh\left(2\zeta/W\right)$, where $\zeta$ is a spatial coordinate along the normal with the origin at the interface.

For ease of implementation, the interfacial surface tension effects can be incorporated within a diffuse interface via a smoothed volumetric force term in a single-field formulation representing the motion of binary fluids. Then, the corresponding single-field incompressible Navier-Stokes equations for binary fluids can be written as
\mathleft
\begin{equation}\label{eqn4}
\qquad \bm{\nabla} \cdot {\bm{u}}=0,
\end{equation}
\mathleft
\begin{equation}\label{eqn5}
\qquad \rho \left( \frac{\partial \bm{u}}{\partial t} + \bm{u}\cdot\bm{\nabla} \bm{u} \right) = - \bm{\nabla} p + \bm{\nabla} \cdot \left[ \mu (\bm{\nabla} \bm{u} + \bm{\nabla} \bm{u}^{\dagger})\right] + \bm{F}_s + \bm{F}_{ext},
\end{equation}
where $\bm{F}_{s}$ is the surface tension force, and $\bm{F}_{ext}$ is the external body force. Here, surface tension force effectively exerts itself in both the normal and tangential directions to the interface as surface tension varies locally with temperature (see Eq.~(\ref{ST_SRF})). The continuous surface force approach~\cite{brackbill1992continuum} is a geometric technique that can be used to account for this, which can be expressed by the following equation involving the Dirac delta function $\delta_s$:
\mathleft
\begin{equation}\label{eqn6}
\qquad \bm{F}_{s} = \left( \sigma \kappa \bm{n} + \bm{\nabla}_{s} \sigma \right)\delta_{s},
\end{equation}
where $\bm{n}$ and $\kappa$ are the unit vector normal and the interface curvature, respectively; they can be obtained from the order parameter via $\bm{n} = \bm{\nabla}{\phi}/|\bm{\nabla}{\phi}|$ and ${\kappa} = \bm{\nabla} \cdot \bm{n}$. In the right side of Eq.~(\ref{eqn6}), the first term is the normal or capillary force acting on the interface, and the second term involves the surface gradient operator $\bm{\nabla}_{s}$ is the tangential or Marangoni force induced by surface tension gradients. Because the surface tension only acts on the interface, the delta function $\delta_{s}$ is required to satisfy $\int_{-\infty}^{+\infty} \delta_{s} dy = 1$. One formulation of $\delta_{s}$, which localizes the smoothed surface tension force suitable within the phase-field modeling framework, is given by $\delta_{s} = 1.5 W |\bm{\nabla} \phi|^2$.

The surface gradient $\bm{\nabla}_{s}$ in Eq.~(\ref{eqn6}) is given by $\bm{\nabla}_{s} = \bm{\nabla} - \bm{n}(\bm{n} \cdot \bm{\nabla})$. Therefore, the Cartesian components of the surface tension force in Eq.~(\ref{eqn6}) can then be expressed as
\mathleft
\begin{eqnarray}
\qquad F_{sx} &=& -\sigma(T) |\bm{\nabla} \phi|^2 (\bm{\nabla} \cdot \bm{n}) {n}_x +|\bm{\nabla} \phi|^2 \left[ ( 1-{n}_x^2 ) \partial_x \sigma(T) - {n}_x {n}_y \partial_y \sigma(T) \right],\nonumber\\
\qquad F_{sy} &=& -\sigma(T) |\bm{\nabla} \phi|^2 (\bm{\nabla} \cdot \bm{n}) {n}_y +|\bm{\nabla} \phi|^2 \left[ ( 1-{n}_y^2 ) \partial_y \sigma(T) - {n}_x {n}_y \partial_x \sigma(T) \right]. \label{eq:surfacetensionforcecomponents}
\end{eqnarray}
Here, the functional dependence of the surface tension on temperature for the SRF is obtained from the nonlinear (parabolic) term that is shown in the equation given in Eq.~(\ref{ST_SRF}). In numerical implementations, in this work, the required spatial gradients $\partial_x \sigma(T) $ and $\partial_y \sigma(T)$ in Eq.~(\ref{eq:surfacetensionforcecomponents}) are calculated using an isotropic finite differencing scheme, which is second order in space but with a beneficial feature of no directional preference~\cite{kumar2004isotropic}. Here, we note that temperature field $T$ is computed by solving the energy transport equation given earlier in Eq.~(\ref{energy eqn}). Finally, the jumps in fluid properties across the interface, such as density and viscosity, can be expressed as a continuous function of the phase field variable, which can then be employed in Eq.~(\ref{eqn5}). We use the following linear interpolation to account for the interfacial variations of fluid properties in this study (see, e.g., \cite{ding2007diffuse}):
\mathleft
\begin{equation}\label{eqn11}
\qquad \rho = \rho_{b} + \frac {\phi - \phi_{b}}{\phi_{a} - \phi_{b}} \left(\rho_{a} - \rho_{b} \right), \quad
\mu = \mu_{b} + \frac {\phi - \phi_{b}}{\phi_{a} - \phi_{b}} \left(\mu_{a} - \mu_{b} \right),
\end{equation}
where $\rho_{a}$, $\rho_{b}$ and $\mu_{a}$, $\mu_{b}$ are the densities and the dynamic viscosities in the fluid phases, respectively and denoted by $\phi_{a}$ and $\phi_{b}$. An equation similar to Eq.~(\ref{eqn11}) will also be utilized for distributing the interfacial jump in the thermal conductivity in solving the energy equation. In this study, we use $\phi_{a}=0$ and $\phi_{b}=1$.

\section{Central moment lattice Boltzmann schemes \label {Sec.LBschemes}}
In this section, we will present a numerical LB approach based on more robust collision models involving central moments~\cite{geier2006cascaded,premnath2009incorporating,premnath2011three,hajabdollahi2021central,yahia2021central} for solving the equations of the phase-field model for tracking the interface (Eq.~(\ref{eqn1})) and the binary fluid motions (Eqs.~(\ref{eqn4})-(\ref{eq:surfacetensionforcecomponents})) given in the previous section, along with transport of energy presented in Eq.~(\ref{energy eqn}) earlier. In general, solving these three equations requires evolving three separate distribution functions on the standard two-dimensional, square lattice (D2Q9) lattice, which involve performing a \emph{collision step} based on the relaxation of different central moments of the distribution function to their equilibria, which is followed by a lock-step advection of the distribution functions to their adjacent nodes along the characteristic directions in the \emph{streaming step}. Then, the macroscopic variables, viz., the order parameter, the fluid pressure and velocity, and the temperature field, are obtained by taking the appropriate lower velocity moments of the respective distribution functions. It should be noted that since the collision step is performed using central moments while the streaming step is performed by means of the distribution functions, this requires the use of appropriate mappings that transform between these quantities pre- and post-collision step. The central moment LB methods are shown to be more robust (e.g., enhanced numerical stability) when compared to the other collision models in the LB framework (see~\cite{hajabdollahi2021central,yahia2021central,yahia2021three} for recent examples). While the recent central moment LB scheme for two-fluid interfacial flows~\cite{hajabdollahi2021central} was constructed using an orthogonal moment basis, in what follows, we will utilize an improved formulation involving an non-orthogonal moment basis. Moreover, for ease of presentation, our discussion will be based on two-dimensional formulations in Cartesian coordinates, on which the axisymmetric effects can be readily introduced as appropriate geometric source terms which will be identified at the end of this section.

\subsection{Interface tracking} \label{Sec.4.1}
We will now discuss a central moment LB technique to solve the conservative ACE given in Eq.~(\ref{eqn1}) by evolving a distribution function $f_\alpha$, where $\alpha=0,1,2,\ldots,8$ represents the discrete particle directions, on the D2Q9 lattice. Generally, during the collision, the set of distribution functions $\mathbf{f}=(f_0,f_1,f_2,\ldots,f_8)^\dagger$ relax to the corresponding equilibrium distribution functions given by $\mathbf{f}^{eq}=(f_0^{eq},f_1^{eq},f_2^{eq},\ldots,f_8^{eq})^\dagger$, which needs to be implemented via their central moments in what follows. Here, and henceforth, the superscript $\dagger$ represents the transpose operator.

In this regard, first, the components of the particle velocities of this lattice can be represented by the following vectors in standard Dirac's bra-ket notation as
\mathleft
\begin{subequations}
\begin{equation} 
\qquad \left| \bm{e}_x \right> = ( 0, 1, 0, -1, 0, 1, -1, -1, 0)^\dagger, \nonumber 
\end{equation}
\begin{equation} 
\qquad \left| \bm{e}_y \right> = ( 0, 0, 1, 0,-1, 1, 1, -1, -1)^\dagger. \nonumber
\end{equation}
\end{subequations}
We also need the following 9-dimensional vector to define the zeroth moment of $f_\alpha$:
\mathleft
\begin{eqnarray} 
\qquad \left|\mathbf{1}\right> = (1,1,1,1,1,1,1,1,1)^{\dag}. \nonumber
\end{eqnarray}
That is, its inner product with the set of distribution functions $\left<\mathbf{f}|\mathbf{1}\right>$ should yield the order parameter $\phi$ of the phase-field model. The central moment LB will then be constructed based on the following set of nine non-orthogonal basis vectors (which differs from the approach presented in~\cite{hajabdollahi2021central}):
\mathleft
\begin{gather}
\qquad \left| P_0 \right> = \left| \mathbf{1} \right>, \quad
\left| P_1 \right> = \left| \bm{e}_x \right>, \quad
\left| P_2 \right> = \left| \bm{e}_y \right>, \nonumber \\[2mm]
\qquad \left| P_3 \right> = \left| \bm{e}_x^2\right>, \quad
\left| P_4 \right> = \left| \bm{e}_y^2 \right>, \quad
\left| P_5 \right> = \left| \bm{e}_x \bm{e}_y \right>,\nonumber \\[2mm]
\qquad \left| P_6 \right> = \left| \bm{e}_x^2 \bm{e}_y \right>,\quad
\left| P_7 \right> = \left| \bm{e}_x \bm{e}_y^2 \right>,\quad
\left| P_8 \right> = \left| \bm{e}_x^2 \bm{e}_y^2 \right>. \nonumber
\end{gather}
Symbols like $ \left| e_x^2 e_y \right> = \left| e_x e_x e_y \right>$ signify a vector that results from the element-wise vector multiplication of vectors $\left| e_x \right>$,$\left| e_x \right>$ and $\left| e_y \right>$. They can be grouped together in the form of the following matrix that maps the distribution functions to the \emph{raw} moments in terms of the above moment basis vectors:
\mathleft
\begin{equation} \label{eqn34}
\qquad \mathbf{P} = \left[
\left|P_0\right>,
\left|P_1\right>,
\left|P_2\right>,
\left|P_3\right>,
\left|P_4\right>,
\left|P_5\right>,
\left|P_6\right>,
\left|P_7\right>,
\left|P_8\right> \right]^\dag.
\end{equation}
Here, it should be noted that the \emph{central} moments are obtained from the distribution moments by shifting the particle velocity $\bm{e}_\alpha$ by the fluid velocity $\bm{u}$. Given these, we can then formally define the raw moments of the distribution function $f_\alpha$ as well as its equilibrium $f_\alpha^{eq}$ as
\mathleft
\begin{subequations}
\begin{equation}
\qquad \left( \begin{array}{c}\kappa'_{mn}\\[2mm] \kappa'^{\;eq}_{mn} \end{array} \right) = \sum_{\alpha = 0}^{8} \left( \begin{array}{c}f_{\alpha} \\[2mm] f_{\alpha}^{eq} \end{array} \right) e_{\alpha x}^m e_{\alpha y}^n,
\end{equation}
and the corresponding central moments as
\mathleft
\begin{equation}
\qquad \left( \begin{array}{c}\kappa_{mn} \\[2mm] \kappa_{mn}^{eq} \end{array} \right) = \sum_{\alpha = 0}^{8} \left( \begin{array}{c}f_{\alpha} \\[2mm] f_{\alpha}^{eq} \end{array} \right) (e_{\alpha x}-u_x)^m ( e_{\alpha y}-u_y)^n.
\end{equation}
\end{subequations}
Thus, $\kappa'_{mn}$ represents the raw moment of order $(m+n)$, while the corresponding central moment is $\kappa_{mn}$. For convenience, we can group all the possible raw moments and the central moments for the D2Q9 lattice via the following two vectors as
\mathleft
\begin{subequations}
\begin{eqnarray}
\qquad \bm{\kappa^{'}} \! \! \! &=& \! \! \! ( \K{00}^{'}, \K{10}^{'},\K{01}^{'}, \K{20}^{'}, \K{02}^{'}, \K{11}^{'},\K{21}^{'}, \K{12}^{'},\K{22}^{'} ),\label{eqn:4a} \\[3mm]
\qquad \bm{\kappa} \! \! \! &=& \! \! \! ( \K{00},\K{10}, \K{01}, \K{20}, \K{02}, \K{11}, \K{21}, \K{12}, \K{22} ).
\end{eqnarray}
\end{subequations}
It should be noted that one can readily map from the distribution functions to the raw moments via $\bm{\kappa^{'}} = \PP\mathbf{f}$, which can then be transformed into the central moments through $\bm{\kappa} = \F \bm{\kappa^{'}}$, where the $\tensr{F}$ follows readily from binomial expansions of $(e_{\alpha x}-u_x)^m ( e_{\alpha y}-u_y)^n$ to relate to $e_{\alpha x}^m e_{\alpha y}^n$ etc. Similarly, the inverse mappings from central moments to raw moments, from which the distribution functions can be recovered via the matrices $\tensr{F}^{-1}$ and $\tensr{P}^{-1}$, respectively. All these mapping relations are explicitly listed in Appendix~\ref{App B}.

As mentioned above, a key aspect of our approach is to perform the collision step such that different central moments shown above relax to their corresponding central moment equilibria. The discrete central moment equilibria $\Keq{mn}$ defined above can be obtained by matching them to the corresponding central moments of the continuous Maxwell distribution function after replacing the density $\rho$ with the order parameter $\phi$; furthermore, the interface sharpening flux terms in the conservative ACE (Eq.~(\ref{eqn1})) need to be accounted for by augmenting the first order central moment equilibrium components with $M_{\phi}\theta n_x$ and $M_{\phi}\theta n_y$~\cite{hajabdollahi2021central}. Thus, we have
\mathleft
\begin{gather}
\qquad \Keq{00} = \phi, \qquad
\Keq{10} = M_{\phi} \theta n_x,\qquad
\Keq{01} = M_{\phi} \theta n_y,\nonumber \\[2mm]
\qquad \Keq{20} = c_{s\phi}^2 \phi,\qquad
\Keq{02} = c_{s\phi}^2 \phi,\qquad
\Keq{11} = 0,\nonumber \\[2mm]
\qquad \Keq{21} = 0,\qquad
\Keq{12} = 0,\qquad
\Keq{22} = c_{s\phi}^4 \phi,
\end{gather}
where $c_{s\phi}^2=1/3$.

Based on the above considerations, inspired by the algorithmic implementation presented in~\cite{geier2015cumulant} (see also~\cite{yahia2021central,yahia2021three}), we can now summarize the central moment LB algorithm for solving the conservative ACE for a time step $\Delta t$ starting from $f_\alpha=f_\alpha(\bm{x},t)$ as follows:
\begin{itemize}
\item Compute pre-collision raw moments from distribution functions via $\bm{\kappa^{'}} = \PP\mathbf{f}$ (see Eq.~(\ref{eq:tensorP}) in Appendix~\ref{App B} for $\tensr{P}$)
\item Compute pre-collision central moments from raw moments via $\bm{\kappa} = \F \bm{\kappa^{'}}$ (see Eq.~(\ref{eq:tensorF}) in Appendix~\ref{App B} for $\tensr{F}$)
\item Perform collision step via relaxation of central moments $\kappa_{mn}$ to their equilibria $\kappa_{mn}^{eq}$: \newline
\begin{equation}\label{eq:centralmomentrelaxationCACE}
\qquad \tilde{\kappa}_{mn} = \kappa_{mn} + \wphi{mn} (\Keq{mn} - \kappa_{mn}),
\end{equation}
where $(mn)=(00),(10),(01),(20),(02),(11),(21),(12)$, and $(22)$, and $\wphi{mn}$ is the relaxation parameter for moment of order ($m+n$). Here, the implicit summation convention of repeated indices is not assumed. The relaxation parameters of the first order moments $\omega_{10}^{\phi} = \omega_{01}^{\phi} = \omega^{\phi}$ are related to the mobility coefficient $M_{\phi}$ in Eq.~(\ref{eqn1}) via $M_{\phi}= c_{s\phi}^2 \left( \frac{1}{\omega^{\phi}} - \frac{1}{2}\right)\Delta t$, and the rest of the relaxation parameters are typically set to unity, i.e., $\wphi{mn}=1.0$, where $(m+n) \geq 2$. The results of Eq.~(\ref{eq:centralmomentrelaxationCACE}) are then grouped in $\bm{\tilde{\kappa}}$.
\item Compute post-collision raw moments from post-collision central moments via $\bm{\tilde{\kappa}^{'}} = \F^{-1} \bm{\tilde{\kappa}}$ (see Eq.~(\ref{eq:tensorFinverse}) in Appendix~\ref{App B} for $\tensr{F}^{-1}$)
\item Compute post-collision distribution functions from post-collision raw moments via $\mathbf{\tilde{f}} = \PP^{-1}\bm{\tilde{\kappa}^{'}}$ (see Eq.~(\ref{eq:tensorPinverse}) in Appendix~\ref{App B} for $\tensr{P}^{-1}$)
\item Perform streaming step via $f_{\alpha}(\bm{x}, t+ \Delta t) = \tilde{f}_{\alpha}(\bm{x}-\bm{e}_{\alpha} \Delta t)$, where $\alpha = 0,1,2,...,8$.
\item Update the order parameter $\phi$ of the phase-field model for interface capturing through \newline
\begin{equation}
\qquad \qquad\phi = \sum_{\alpha=0}^{8} f_{\alpha}.
\end{equation}
\end{itemize}
\subsection{Two-fluid motions}\label{Sec.4.2}
Next, we will present a central moment LB scheme to solve the motion of binary fluids with interfacial forces represented in Eqs.~(\ref{eqn4})-(\ref{eq:surfacetensionforcecomponents}) by evolving another distribution function $g_\alpha$, where $\alpha=0,1,2,\ldots,8$. Our approach is based on a discretization of the modified continuous Boltzmann equation and obtaining the discrete central moment equilibria and central moments of the source terms for the body forces via a matching principle with their continuous counterparts as detailed in Ref.~\cite{hajabdollahi2021central}. However, in contrast to Ref.~\cite{hajabdollahi2021central}, where an orthogonal moment basis is employed, resulting in the so-called cascaded LB approach, in the following, we consider the simpler, non-orthogonal moment basis vectors as given earlier in Eq.~(\ref{eqn34}).

As in the previous section, we first define the following raw moments and the central moments of the distribution function $g_\alpha$, its equilibrium $g_\alpha^{eq}$, as well as the source term $S_\alpha$, where the latter accounts for the surface tension and body forces, as well as those that arise from the application of a transformation to simulate flows at high-density ratios in the incompressible limit (see~\cite{he1999lattice,hajabdollahi2021central}):
\mathleft
\begin{subequations}
\begin{equation}
\qquad \left( \begin{array}{c} {\eta}'_{ mn}\\[1mm] \eta'^{\;eq}_{ mn}\\[1mm] {\sigma}'_{ mn} \end{array} \right) = \sum_{\alpha = 0}^{8} \left( \begin{array}{c}g_{\alpha} \\[1mm] g_{\alpha}^{eq}\\[1mm] S_\alpha \end{array} \right) e_{\alpha x}^m e_{\alpha y}^n,
\end{equation}
\mathleft
\begin{equation}
\qquad \left( \begin{array}{c}{\eta}_{ mn}\\[1mm] \eta^{\;eq}_{ mn}\\[1mm] {\sigma}_{ mn} \end{array} \right) = \sum_{\alpha = 0}^{8} \left( \begin{array}{c}g_{\alpha} \\[1mm] g_{\alpha}^{eq}\\[1mm] S_\alpha \end{array} \right) (e_{\alpha x}-u_x)^m ( e_{\alpha y}-u_y)^n.
\end{equation}
\end{subequations}
For convenience, we can group the elements of the distribution function, its equilibrium, and the source term for the D2Q9 lattice as the following vectors: $\mathbf{g}=(g_0,g_1,g_2,\ldots,g_8)^\dagger$, $\mathbf{g}^{eq}=(g_0^{eq},g_1^{eq},g_2^{eq},\ldots,g_8^{eq})^\dagger$, and $\mathbf{S}=(S_0,S_1,S_2,\ldots,S_8)^\dagger$. Moreover, we group all the possible raw moments and the central moments defined above for the D2Q9 lattice via the following:
\mathleft
\begin{subequations}
\begin{eqnarray}
\qquad \bm{{\eta}^{'}} \! \! \! &=& \! \! \! ( {\eta}_{00}^{'}, {\eta}_{10}^{'},{\eta}_{01}^{'}, {\eta}_{20}^{'}, {\eta}_{02}^{'}, {\eta}_{11}^{'},{\eta}_{21}^{'}, {\eta}_{12}^{'},{\eta}_{22}^{'} ),\label{eqn:4a} \\[3mm]
\qquad \bm{{\eta}} \! \! \! &=& \! \! \! ( {\eta}_{00},{\eta}_{10}, {\eta}_{01}, {\eta}_{20}, {\eta}_{02}, {\eta}_{11}, {\eta}_{21}, {\eta}_{12}, {\eta}_{22} ),
\end{eqnarray}
\end{subequations}
and similarly for raw moments and the central moments the equilibrium and the source term.

The collision step will be performed such that different central moments shown above relax to their corresponding central moment equilibria, which are augmented by changes in the central moments due to the net forces; the latter is given by sum the surface tension force $\bm{F}_s=(F_{sx},F_{sy})$, which can have contributions from both the capillary and Marangoni forces as represented in Eq.~(\ref{eq:surfacetensionforcecomponents}), and any external force $\bm{F}_{ext}=(F_{ext,x},F_{ext,y})$, i.e., $\bm{F}_{t}=\bm{F}_{s}+\bm{F}_{ext}$ or $(F_{tx},F_{ty})=(F_{sx}+F_{ext,x},F_{sy}+F_{ext,y})$. Moreover, the use of an incompressible transformation, as mentioned above, leads to a pressure-based formulation, involving the incorporation of a net pressure force $\bm{F}_p$ arising from $\varphi(\rho)=p-\rho c_s^2$, i.e., $\bm{F}_p=-\bm{\nabla}\varphi$, or $(F_{px},F_{py})=(-\partial_x\varphi,-\partial_y\varphi)$ (see~\cite{hajabdollahi2021central} for details). Then, the discrete central moment equilibria $\eta_{mn}$ defined above can be obtained by matching them to the corresponding continuous central moments of the equilibrium that arise from the incompressible transformation, and similarly for the central moments of the source term $\sigma_{mn}$, which then results in the following expressions for the D2Q9 lattice~\cite{hajabdollahi2021central}:
\mathleft
\begin{gather}
\qquad {\eta}_{00}^{eq} = p, \quad {\eta}_{10}^{eq} = -\varphi(\rho) u_x, \quad {\eta}_{01}^{eq} = -\varphi(\rho)u_y, \quad {\eta}_{20}^{eq} = p c_s^2 + \varphi(\rho)u_x^2,\nonumber \\
\qquad {\eta}_{02}^{eq} = p c_s^2 + \varphi(\rho)u_y^2, \quad {\eta}_{11}^{eq} = \varphi(\rho)u_x u_y , \quad {\eta}_{21}^{eq} = -\varphi(\rho)(u_x^2+ c_s^2) u_y,\nonumber \\
\qquad {\eta}_{12}^{eq} = -\varphi(\rho)(u_y^2+ c_s^2) u_x, \quad {\eta}_{22}^{eq} = c_s^6 \rho + \varphi(\rho)(u_x^2+ c_s^2) (u_y^2+ c_s^2).
\end{gather}
and
\begin{gather}
\qquad {\sigma}_{00}= \Gamma_{00}^p, \quad {\sigma}_{10} = c_s^2 F_{tx}-u_x{\Gamma}_{00}^p, \quad {\sigma}_{01} = c_s^2 F_{ty}-u_y{\Gamma}_{00}^p, \nonumber \\
\qquad {\sigma}_{20} = 2c_s^2 F_{px}u_x+(u_x^2+c_s^2){\Gamma}_{00}^p,\quad {\sigma}_{02} = 2c_s^2 F_{py}u_y+(u_y^2+c_s^2){\Gamma}_{00}^p, \nonumber\\
\qquad {\sigma}_{11} = c_s^2 (F_{px}u_y+F_{py}u_x)+u_x u_y{\Gamma}_{00}^p,\quad {\sigma}_{21} = 0, \quad {\sigma}_{12} = 0, \quad {\sigma}_{22} = 0,
\end{gather}
where $\Gamma_{00}^p=(F_{px}u_x+F_{py}u_y)$.

Using the above developments, we can now summarize the central moment LB algorithm for computing the two-fluid motion with interfacial forces for a time step $\Delta t$ starting from $g_\alpha=g_\alpha(\bm{x},t)$ as follows:
\begin{itemize}
\item Compute pre-collision raw moments from distribution functions via $\bm{\eta^{'}} = \PP\mathbf{g}$ (see Eq.~(\ref{eq:tensorP}) in Appendix~\ref{App B} for $\tensr{P}$)
\item Compute pre-collision central moments from raw moments via $\bm{\eta} = \F \bm{\eta^{'}}$ (see Eq.~(\ref{eq:tensorF}) in Appendix~\ref{App B} for $\tensr{F}$)
\item Perform collision step via relaxation of central moments $\eta_{mn}$ to their equilibria $\eta_{mn}^{eq}$ and augmented with the source terms $\sigma_{mn}$: \newline
In order to allow for an independent specification of the shear viscosity $\nu$ from the bulk viscosity $\zeta$, the trace of the second order moments ${\eta}_{20} + {\eta}_{02}$ should be evolved independently from the other second-order moments. To accomplish this, prior to performing collision, we combine the diagonal parts of the second-order moments as follows (see, e.g.,~\cite{geier2015cumulant,yahia2021central,yahia2021three}):
\mathleft
\begin{gather}
\qquad {\eta}_{2s} = {\eta}_{20} + {\eta}_{02}, \qquad {\eta}_{2s}^{eg} = {\eta}_{20s}^{eg} + {\eta}_{02}^{eg}, \qquad {\sigma}_{2s} = {\sigma}_{20s} + {\sigma}_{02},\nonumber \\[2mm]
\qquad {\eta}_{2d} = {\eta}_{20} - {\eta}_{02}, \qquad {\eta}_{2d}^{eg} = {\eta}_{20s}^{eg} - {\eta}_{02}^{eg}, \qquad {\sigma}_{2d} = {\sigma}_{20s} - {\sigma}_{02},\nonumber
\end{gather}
and thus ${\eta}_{2s}$ and ${\eta}_{2d}$ will be evolved independently under collision. Then, the post-collision central moments under relaxation and augmentation due to the forces can be computed via
\mathleft
\begin{equation} \label{eq:centralmomentrelaxationtwofluidmotion}
\qquad \tilde{\eta}_{ mn} = {\eta}_{mn} + \omega_{mn} \left( {\eta}_{mn}^{eq} -{\eta}_{mn} \right) + \left(1-\omega_{mn}/2 \right) {\sigma}_{mn}\Delta t,
\end{equation}
where $\omega_{mn}$ is the relaxation time corresponding to the central moment ${\eta}_{ mn}$, and $(mn)= (00), (10), (01), (2s), (2d), (11), (21), (12), \mbox{and}, (22)$. Here, the relaxation parameter $\omega_{2s}$ is related to the bulk viscosity via $\zeta= c_s^2 \left(1/\omega_{2s}- 1/2\right)\Delta t$, while the relaxation parameters $\omega_{2d}$ and $\omega_{11}$ are related to shear viscosity via $\nu= c_s^2 \left( 1/\omega_{ij} - 1/2\right)\Delta t$ where $(ij)=(2d),(11)$. Typically, $c_s^2=1/3$. Given Eq.~(\ref{eqn11}), it should be noted that if the bulk fluid properties are different, the relaxation parameters $\omega_{2d}$ and $\omega_{11}$ will then vary locally across the interface. The rest of the relaxation parameters of central moments are generally set to unity, i.e., $\omega_{ij}=1.0$, where $(ij)=(00),(10),(01),(2s),(21),(12),(22)$.\newline
Also, the combined forms of the post-collision central moments $\tilde{\eta}_{2s}$ and $\tilde{\eta}_{2d}$ are then segregated in their individual components $\tilde{\eta}_{20}$ and $\tilde{\eta}_{02}$ via
\mathleft
\begin{equation*}
\qquad \tilde{\eta}_{20} = \frac{1}{2} \left(\tilde{\eta}_{2s}+\tilde{\eta}_{2d} \right), \qquad \tilde{\eta}_{02} = \frac{1}{2} \left(\tilde{\eta}_{2s}-\tilde{\eta}_{2d} \right).
\end{equation*}
Finally, the results of Eq.~(\ref{eq:centralmomentrelaxationtwofluidmotion}) by accounting for the above segregation are then grouped in $\bm{\tilde{\eta}}$.
\item Compute post-collision raw moments from post-collision central moments via $\bm{\tilde{\eta}^{'}} = \F^{-1} \bm{\tilde{\eta}}$ (see Eq.~(\ref{eq:tensorFinverse}) in Appendix~\ref{App B} for $\tensr{F}^{-1}$)
\item Compute post-collision distribution functions from post-collision raw moments via $\mathbf{\tilde{g}} = \PP^{-1}\bm{\tilde{\eta}^{'}}$ (see Eq.~(\ref{eq:tensorPinverse}) in Appendix~\ref{App B} for $\tensr{P}^{-1}$)
\item Perform streaming step via $g_{\alpha}(\bm{x}, t+ \Delta t) = \tilde{g}_{\alpha}(\bm{x}-\bm{e}_{\alpha} \Delta t)$, where $\alpha = 0,1,2,...,8$.
\item Update the pressure field $p$ and the components of the fluid velocity $\bm{u}=(u_x,u_y)$ through \newline
\begin{equation}
\qquad \qquad p = \sum_\alpha g_\alpha + \frac{1}{2} \bm{F}_p \cdot {\bm{u}}\Delta t,\quad \rho c_s^2 \bm{u} = \sum_\alpha g_\alpha \bm{e}_\alpha + \frac{1}{2} c_s^2 \bm{F}_{t}\Delta t.
\end{equation}
\end{itemize}

\subsection{Energy transport} \label{Sec.4.3}
We will now discuss a central moment LB approach for the solution of the energy transport equation (Eq.~(\ref{energy eqn})) by evolving a third distribution function $h_\alpha$, where $\alpha=0,1,2,\ldots,8$, on the D2Q9 lattice. Since Eq.~(\ref{energy eqn}) is an advection-diffusion equation, its construction procedure is quite similar to that of the LB scheme for the conservative ACE presented earlier, albeit without the presence of a term such as the interface sharpening flux term which appears in the latter case.
As before, we first define the following raw moments and central moments, respectively, of the distribution function $h_\alpha$, as well as its equilibrium $h_\alpha^{eq}$:
\mathleft
\begin{subequations}
\begin{equation}
\qquad \left( \begin{array}{c}\chi'_{mn}\\[2mm] \chi'^{\;eq}_{mn} \end{array} \right) = \sum_{\alpha = 0}^{8} \left( \begin{array}{c} h_{\alpha} \\[2mm] h_{\alpha}^{eq} \end{array} \right) e_{\alpha x}^m e_{\alpha y}^n,
\end{equation}
\mathleft
\begin{equation}
\qquad \left( \begin{array}{c}\chi_{mn} \\[2mm] \chi_{mn}^{eq} \end{array} \right) = \sum_{\alpha = 0}^{8} \left( \begin{array}{c} h_{\alpha} \\[2mm] h_{\alpha}^{eq} \end{array} \right) (e_{\alpha x}-u_x)^m ( e_{\alpha y}-u_y)^n.
\end{equation}
\end{subequations}
For convenience, we list the components of the distribution function and its equilibrium, respectively, using $\mathbf{h}=(h_0,h_1,h_2,\ldots,h_8)^\dagger$ and $\mathbf{h}^{eq}=(h_0^{eq},h_1^{eq},h_2^{eq},\ldots,h_8^{eq})^\dagger$, and analogously for the raw moments and central moments via
\mathleft
\begin{subequations}
\begin{eqnarray}
\qquad \bm{\chi^{'}} \! \! \! &=& \! \! \! ( \chi_{00}^{'}, \chi_{10}^{'},\chi_{01}^{'}, \chi_{20}^{'}, \chi_{02}^{'}, \chi_{11}^{'},\chi_{21}^{'}, \chi_{12}^{'},\chi_{22}^{'} ),\label{eqn:4a} \\[3mm]
\qquad \bm{\chi} \! \! \! &=& \! \! \! ( \chi_{00},\chi_{10}, \chi_{01}, \chi_{20}, \chi_{02}, \chi_{11}, \chi_{21}, \chi_{12}, \chi_{22} ).
\end{eqnarray}
\end{subequations}
To construct a central moment-based collision model for solving the energy equation, similar to Sec.~\ref{Sec.4.1}, we obtain the discrete equilibrium central moments from the corresponding continuous counterpart of the Maxwellian by replacing the density $\rho$ with the temperature $T$, and the results read as
\mathleft
\begin{gather}
\qquad \chi_{00}^{eq} = T, \qquad
\chi_{10}^{eq} = 0,\qquad
\chi_{01}^{eq} = 0,\nonumber \\[2mm]
\qquad \chi_{20}^{eq} = c_{sT}^2 T,\qquad
\chi_{02}^{eq} = c_{sT}^2 T,\qquad
\chi_{11}^{eq} = 0,\nonumber \\[2mm]
\qquad \chi_{21} = 0,\qquad
\chi_{12}^{eq} = 0,\qquad
\chi_{22}^{eq} = c_{sT}^4 T,
\end{gather}
where, typically, $c_{sT}^2 = 1/3$. Then, the computational procedure for solving the energy equation for a time step $\Delta t$ starting from $h_\alpha=h_\alpha(\bm{x},t)$ can be summarized as follows:
\begin{itemize}
\item Compute pre-collision raw moments from distribution functions via $\bm{\chi^{'}} = \PP\mathbf{h}$ (see Eq.~(\ref{eq:tensorP}) in Appendix~\ref{App B} for $\tensr{P}$)
\item Compute pre-collision central moments from raw moments via $\bm{\chi} = \F \bm{\chi^{'}}$ (see Eq.~(\ref{eq:tensorF}) in Appendix~\ref{App B} for $\tensr{F}$)
\item Perform collision step via relaxation of central moments $\chi_{mn}$ to their equilibria $\chi_{mn}^{eq}$: \newline
\begin{equation}\label{eq:centralmomentrelaxationenergyequation}
\qquad \tilde{\chi}_{mn} = \chi_{mn} + \omega^T_{mn} (\chi_{mn}^{eq} - \chi_{mn}),
\end{equation}
where $(mn)=(00),(10),(01),(20),(02),(11),(21),(12)$, and $(22)$, and $\omega^T_{mn}$ is the relaxation parameter for moment of order ($m+n$). The relaxation parameters of the first order moments $\omega_{10}^T$=$\omega_{01}^T$=$ \omega^T$ are related to the thermal diffusivity $\alpha=k/(\rho c_p)$ via $\alpha = c_{sT}^2 \left(1/\omega^{T} - 1/2\right)\Delta t$, and the rest of the relaxation parameters of higher central moments are typically set to unity. The results of Eq.~(\ref{eq:centralmomentrelaxationenergyequation}) are then grouped in $\bm{\tilde{\chi}}$.
\item Compute post-collision raw moments from post-collision central moments via $\bm{\tilde{\chi}^{'}} = \F^{-1} \bm{\tilde{\chi}}$ (see Eq.~(\ref{eq:tensorFinverse}) in Appendix~\ref{App B} for $\tensr{F}^{-1}$)
\item Compute post-collision distribution functions from post-collision raw moments via $\mathbf{\tilde{h}} = \PP^{-1}\bm{\tilde{\chi}^{'}}$ (see Eq.~(\ref{eq:tensorPinverse}) in Appendix~\ref{App B} for $\tensr{P}^{-1}$)
\item Perform streaming step via $h_{\alpha}(\bm{x}, t+ \Delta t) = \tilde{h}_{\alpha}(\bm{x}-\bm{e}_{\alpha} \Delta t)$, where $\alpha = 0,1,2,...,8$.
\item Update the temperature field $T$ is obtained from \newline
\begin{equation}
\qquad \qquad T = \sum_{\alpha=0}^{8} h_{\alpha}.
\end{equation}
\end{itemize}

\subsection{Axisymmetric computational modeling} \label{DOLI HF Axisymmetric Model}
In order to incorporate three-dimensional effects existing in axisymmetric multiphase thermal flows in a two-dimensional framework, the numerical methods based on the LBM on the standard two-dimensional, square lattice (D2Q9) for interface tracking, two fluid motions, and energy transport discussed above requires some further extensions. Our approach is based on using source terms to include three-dimensional axisymmetric effects in a quasi two-dimensional formulation, which was first introduced in~\cite{premnath2005lattice} and then was later extended and applied by many other researchers. In this work, we adopt such a strategy using a robust central moments based LB formulation, which has been tested and validated for different benchmark problems recently~\cite{elbousefi2023axisymmetric}.

In essence, this approach involving expressing the conservative Allen-Cahn equation (CACE), two-fluid mass and momentum equations, and the energy equation first in the cylindrical coordinates with axial symmetry and then by applying a coordinate transformation and expressing the resulting equations in the two-dimensional Cartesian coordinates which effectively involves additional contributions related to geometric source terms. Such a strategy is modular in construction in that they can be introduced as simple modifications to the existing two-dimensional implementations such as the central moment LBMs discussed previously. In the following, by applying a coordinate transformation $(z,r) \rightarrow (x,y)$ to the axisymmetric CACE, equations of the two-fluid motions, and the energy transport fequation, we will identify the relevant source term contributions. As such, the incorporation of the source terms in the LB schemes are relatively straightforward and the details are omitted here for brevity.

The axisymmetric CACE with an additional source term rewritten in an effective $2D$ Cartesian coordinate system is as follows.
\mathleft
\begin{equation} \label{Axi_3}
\qquad \partial\sbs{t}\phi+\partial\sbs{j}(\phi u\sbs{j})=\partial\sbs{j} \left[ M\sbs{\phi} \left(\partial\sbs{j}\phi - \theta n\sbs{j} \right) \right] - u\sbs{y} \phi/y.
\end{equation}
Here, except for the treatment of the convective flux terms, the diffusion term, which is a numerical artifact in interface tracking, is still carried out in the Cartesian coordinates to maintain its mass conservation property. This is due to the fact at the macroscopic scales involved in the multiphase flow simulations, the interface thickness is an artificial parameter and the stabilizing diffusion term is counteracted by an interface sharpening term in a numerical sense in the `sharp' interface limit~\cite{sun2007sharp}, and in fact the CACE can be regarded as being analogous to the conservative level-set method which was constructed solely based on numerical considerations for interface tracking~\cite{olsson2005conservative}. Moreover, in the LB context, a similar treatment to handle the diffusion term as mentioned above was considered in a recent work in~\cite{xu2022modified} to maintain mass conservation in axisymmetric simulations. Thus, effectively, the required source term for the LB solver for interface tracking can be identified as $S_\phi = - u\sbs{y} \phi/y$. Next, in the 2D Cartesian coordinate system, the axisymmetric continuity equation for the two-fluid motions recovered by the LB approach is
\mathleft
\begin{equation} \label{Axi_1}
\qquad \partial\sbs{t} p+\partial\sbs{i} \left( \rho c_s^2 u\sbs{i} \right) = u\sbs{i} F_{p,i} - c_s^2\rho u\sbs{y}/y.
\end{equation}
where the last term of Eq.~(\ref{Axi_1}) represents the geometric source term, i.e., $S_m = - c_s^2\rho u\sbs{y}/y$. Here, the pressure force can be presented as $F_{p,i}=-\partial\sbs{i}\left(p - c_s^2 \rho\right)$. Moreover, the axisymmetric two-fluid momentum conservation equation recovered by the LBM in a single-field formulation in the 2D Cartesian coordinates reads as
\mathleft
\begin{equation} \label{Axi_2}
\begin{split}
\qquad \partial\sbs{t}(\rho u\sbs{i}) +\partial\sbs{j}( \rho u\sbs{j} u\sbs{i}) = &-\partial\sbs{i}p + \partial\sbs{i}\left[\rho \nu \left(\partial\sbs{j} u\sbs{i} +\partial\sbs{i} u\sbs{j}\right)-\rho \nu \partial\sbs{k} u\sbs{k} \delta\sbs{ij}+\rho \zeta \partial\sbs{k} u\sbs{k} \delta\sbs{ij} \right] + \\
&+\partial\sbs{i} \left[ \rho \nu \left( 1- \frac{2}{3} \right) \partial\sbs{k} u\sbs{k} + \rho \left(\zeta-\frac{2}{3} \nu \right) \frac{u\sbs{y}}{y}\right]+ F_{t,i}.
\end{split}
\end{equation}
This equation (Eq.~(\ref{Axi_2})) contains additional geometric force terms appearing in the net total force term $F_{t,i}$ (see below) and an additional term that accounts for the correct shear and bulk viscous effects in axisymmetric flows; the former will be incorporated as source terms, while the latter via an extended moment equilibrium term in the LBM~\cite{elbousefi2023axisymmetric}.
The total force $F_{t,i}$ in above equation can be explicitly written as
\mathleft
\begin{equation} \label{}
\qquad F_{t,i}= F_{s,i} + F_{ext,i}-\rho u\sbs{i} u\sbs{y}/y + \rho \nu \left(\partial\sbs{i} u\sbs{y} + \partial\sbs{y} u\sbs{i} \right)/y - 2 \rho \nu u\sbs{y}\delta\sbs{iy}/y,
\end{equation}
where $F_{s,i}$ and $F_{ext,i}$ are the surface tension force and external body force (e.g., gravity), respectively, which were already discussed in the previous section, and the rest of the equation represents the geometric source terms, which can be identified separately as $F_{geo, i} = -\rho u\sbs{i} u\sbs{y}/y + \rho \nu \left(\partial\sbs{i} u\sbs{y} + \partial\sbs{y} u\sbs{i} \right)/y - 2 \rho \nu u\sbs{y}\delta\sbs{iy}/y$ that arise due to the axisymmetric contributions to the two-dimensional formulation. Finally, the axisymmetric energy equation in 2D Cartesian coordinates is also obtained by applying the above coordinate transformation, which then yields
\mathleft
\begin{equation}
\qquad \partial\sbs{t}T+\partial\sbs{j}(u\sbs{j}T)=\partial\sbs{j}(\alpha \partial\sbs{j}T)+ \alpha \partial\sbs{y}T/y-u\sbs{y}T/y,         \label{Axi_4}
\end{equation}
where the last two terms represent the required geometric source terms: $S_T = \alpha \partial\sbs{y}T/y-u\sbs{y}T/y$. We end this section by noting that the various source terms $S_\phi$, $S_m$, $F_{geo,i}$, and $S_T$ identified above can be used to augment the collision steps of the respective LB solvers via the trapezoidal rule with additional contributions in the lower moment calculations to obtain the field variables such as $\phi$, $p$, $\bm{u}$, and $T$~\cite{elbousefi2023axisymmetric}. The numerical methods as discussed above were implemented using the C++ computer programming language and simulations were performed using a standard Dell computer workstation.

\section{Numerical validation: Interaction of an axisymmetric drop with the interface of an unheated liquid layer} \label{DOLI Benchmark}
The two-dimensional central moment LB schemes for multiphase flows have been extensively validated in~\cite{hajabdollahi2021central} for a variety of standard benchmark cases, which was further tested and studied for problems related to thermocapillary effects, including Marangoni effects in our previous investigations~\cite{elbousefi2023thermocapillary,elbousefi2024investigation}. In particular, in our recent work~\cite{elbousefi2023thermocapillary}, we demonstrated the accuracy of the LB formulation discussed in Sec.~\ref{Sec.LBschemes} for simulation thermocapillary convection of SRF layers by comparisons against a new analytical solution developed for this purpose; moreover, we further validated and used it to study problems related to bubble migration in SRFs and various capillary phenomena including the Laplace pressure test in~\cite{elbousefi2024investigation}. In addition, the LB implementation of the axisymmetric extension discussed in the previous section was validated against various standard multiphase flow benchmarks in~\cite{elbousefi2023axisymmetric}. Nevertheless, we further test our approach here for a specific configuration that is directly relevant to the subject of our present investigation, viz., the partial coalescence or pinch-off of an axisymmetric liquid drop interacting with the interface of an unheated liquid layer, which has been studied in detail both experimentally and numerically in the case of normal fluids (NFs) in~\cite{blanchette2006partial}, which we use as a benchmark problem for validation purpose.

In our simulation setup in which the fluids are initially set with a zero velocity, a liquid drop of radius $R$, is initially placed above a flat liquid interface, which causes a thin `neck' or connection to form between the drop and the interface. This neck serves as the starting point for the coalescence process, which is further modulated during the later stages depending on the relative strengths of the various competing forces. The computational domain is resolved in the vertical and horizontal directions using a grid resolution of $12R\times12R$, where $R=50$; the flat interface and the drop are initially placed at the distances of $2R$ and $3R$, respectively, from the bottom side, which is based on the set up reported in a previous study~\cite{blanchette2006partial} (see Fig.~\ref{DOLI_Initial_Condition} for a schematic arrangement). As is standard for LB simulations, we use lattice units. The no-slip boundary conditions on the walls are specified in the LB scheme using the standard half-way bounce back approach~\cite{ladd1994numerical}, while the no-gradient condition along the axis boundary is enforced via the half-way mirror or free-slip condition of the particular distribution functions (see e.g.,~\cite{kruger2017lattice}).

To characterize the strength of viscous forces relative to the surface tension forces, we specify the dimensionless Ohnesorge number; in addition, the relative effects of the gravity force with respect to the surface tension force is given in terms of the Bond number (see Sec.~\ref{DOLI HF Subsection.1.2} for their definitions). The outcome of the interaction between the drop and the initially flat liquid layer is either coalescence (or no pinch-off)  and pinch-off, which is dictated by the relative magnitudes of the Ohnesorge number $\mbox{Oh}$ and the Bond number $\mbox{Bo}$ for chosen density and viscosity ratios, i.e., $\tilde{\rho}$ and $\tilde{\mu}$. It should be pointed out that the three-dimensional effects as represented by the axisymmetric formulation is critical to reproduce the dynamics and the outcomes of this case study accurately; this is due to the fact that a purely two-dimensional formulation is insufficient to model the necessary details in a physically realistic manner. Hence, this problem is serves as a good test case to investigate the efficacy of our approach. Reference~\cite{blanchette2006partial} reported a regime map of the pinch-off or no-pinch-off as a $\mbox{Oh}$-$\mbox{Bo}$ diagram at $\tilde{\rho} = 15.72$ and $\tilde{\mu} = 21$ obtained from observations from an experiment as well as a $\mbox{Oh}$-$\mbox{Bo}$ curve that delineates the two different outcomes obtained from a numerical simulation, with the latter having been shown to be in good quantitative agreement with the former.

\begin{figure}[H]
\centering
\includegraphics[trim = 0 0 0 0, clip, width =90mm]{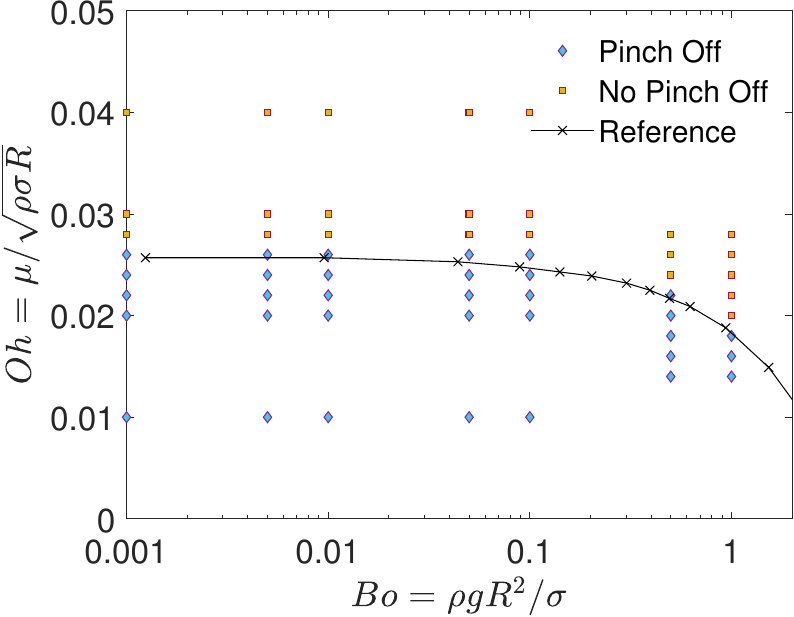}
\caption{Comparison of critical Ohnesorge number as a function of the Bond number obtained from our axisymmetric multiphase central moment LBM simulations (symbols) with the numerical data from Blanchette and Bigioni (2006) (lines)~\cite{blanchette2006partial} for $\tilde{\rho} = 15.72$ and $\tilde{\mu} = 21$.}
\label{DOLI_benchmark}
\end{figure}

We performed simulations of this set up using our axisymmetric central moment LB approach for various sets of $\mbox{Bo}$ and $\mbox{Oh}$. The CACE model parameters $M_{\phi}$ and $W$ are chosen as $0.1$ and $5$, respectively. The pinch-off outcome is recorded by a blue symbol and the no-pinch-off scenario by a red symbol and these results are plotted in Fig.~\ref{DOLI_benchmark}. Also, shown in this figure is the curve for the critical $\mbox{Oh}$ as a function of $\mbox{Bo}$ that delineates the two possible outcomes based on the reference data~\cite{blanchette2006partial}. It can be seen that our simulation results are in very good agreement with the benchmark data confirming the validity of our numerical approach for studying the coalescence/pinch-off processes during the interaction of a drop with a liquid layer.

\section{Results and discussion: Imposed heat flux boundary condition case} \label{DOLI HF Simulation Results and Discussion}
We now study the interfacial dynamics and thermocapillary flows during a self-rewetting drop interacting with a nonuniformly heated fluid layer by an imposed heat flux at its bottom side (Fig.~\ref{HFBC_heated_MODEL}) in order to examine the effects of the Ohnesorge number $\mbox{Oh}$ and the Bond number $\mbox{Bo}$. To achieve this, the LB schemes, which were verified for a related canonical problem in the previous section, are employed. The study begins by focusing on scenarios where the dimensionless surface tension quadratic sensitivity coefficient, represented as $\sigma_{TT}$ or $M_2$ (see Eq.~(\ref{three-prime}) for definitions), is not zero. This is done to showcase the influence of self-rewetting fluids (SRFs) and to compare the outcomes with those of normal fluids (NFs) where only the linear coefficient of surface tension $\sigma_{T}$ (or $M_1$) is present. As such, the strength of thermocapillary flows is determined by selecting two dimensionless coefficients: the linear coefficient  $M_1$, and the quadratic coefficient $M_2$, which describe how surface tension changes with temperature. Unless otherwise stated, the following parameters are maintained at constant values throughout this study: $R=50$, $\sigma_o = 0.005$, $\tilde{\rho} = 1000$, $\tilde{\mu} = 100$, and $\tilde{\alpha} = 1$. For the model parameters in the conservative ACE for interface tracking, we chose the interface thickness and the mobility as $W = 5$ and $M_{\phi} = 0.1$, respectively.

\subsection{Partial and complete coalescence of a non-heated drop on a fluid layer} \label{DOLI HF P and C coalescence }
We will first explain the general coalescence/pinch-off mechanism of a drop on a liquid pool before presenting our results. When there is a significant delay in the vertical collapse due to the presence of converging capillary waves, it allows the subsequent horizontal collapse to effectively pinch the neck off the structure, resulting in the formation of a smaller droplet, often referred to as a daughter droplet. Conversely, depending on the interplay between the viscous and surface tension forces, suppose that these waves are effectively suppressed before reaching their convergence point at the peak, preventing the drop from stretching adequately. The drop then will undergo complete coalescence with the surrounding fluid in that case. The line that separates these two outcomes, partial and complete coalescence, is defined by the critical Ohnesorge number, $\mbox{Oh}^{\star}$. Importantly, this transition point is only minimally influenced by the Bond number, showcasing a subtle interplay between these two parameters.

The relative magnitudes of $\mbox{Bo}$ and $\mbox{Oh}$ determine the relative roles of various competing forces that control the processes associated with the drop's coalescence. A drop may coalesce completely or partially, depending on the strength of surface tension, gravity, and viscous forces. The relatively small length and time scales that result from the multi-scale nature of this phenomenon make the simulation of partial coalescence particularly challenging. In this example, we consider a liquid droplet with a radius of $R=50$. Initially, the droplet is positioned above a flat interface within a computational domain with a resolution $12R\times12R$. The initial placement of the flat interface is at a distance of $2R$ below the axis, while the droplet is initially located at a distance of $3R$ from the same axis. The configuration is designed so that the boundaries, much like in the reference~\cite{constante2021role}, are distant from the droplet and do not significantly influence the coalescence process's physical behavior. The no-slip and free-slip conditions for the solid wall and axis boundaries, respectively, are implemented in the LB schemes via the half-way bounce-back and the mirroring of the distribution functions, respectively, as pointed out in the previous section. In addition, to account for thermal effects, the imposed temperature boundary conditions are specified via the so-called anti-bounce back approach, while the heat flux using a bounce back method augmented with a source term for the magnitude of the local heat flux~\cite{kruger2017lattice}.

\begin{figure}[H]
\centering
\begin{subfigure}{0.32\textwidth}
\includegraphics[trim = 0 0 0 0, clip, width =47mm]{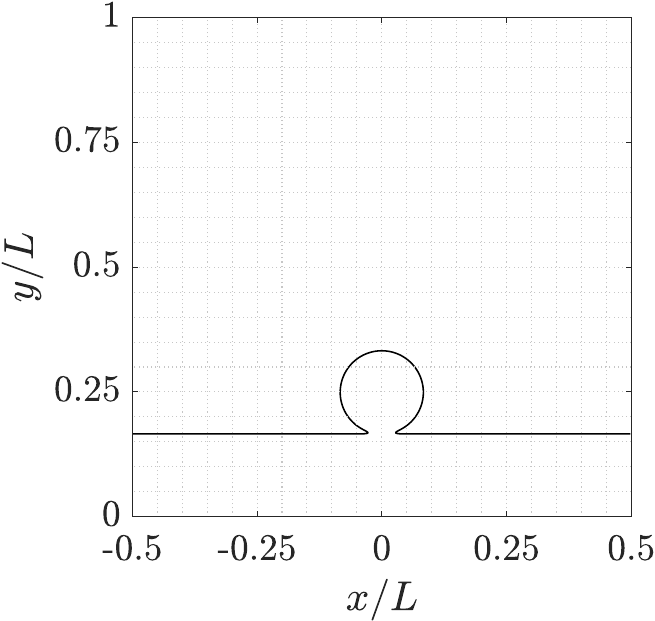}
\caption{\quad \quad $t/T\sbs{o} = 0$}
\label{initial_condition1}
\end{subfigure}
\begin{subfigure}{0.32\textwidth}
\includegraphics[trim = 0 0 0 0, clip, width =47mm]{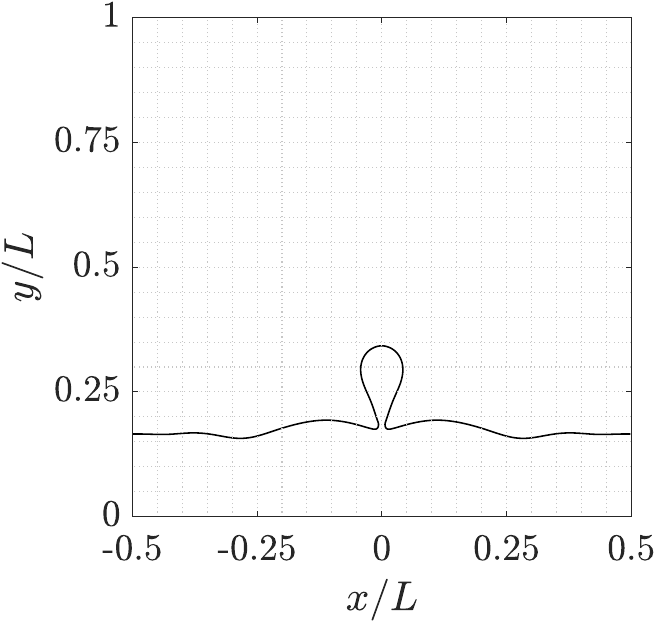}
\caption*{$t/T\sbs{o} = 1.65$}
\end{subfigure}
\begin{subfigure}{0.32\textwidth}
\includegraphics[trim = 0 0 0 0, clip, width =47mm]{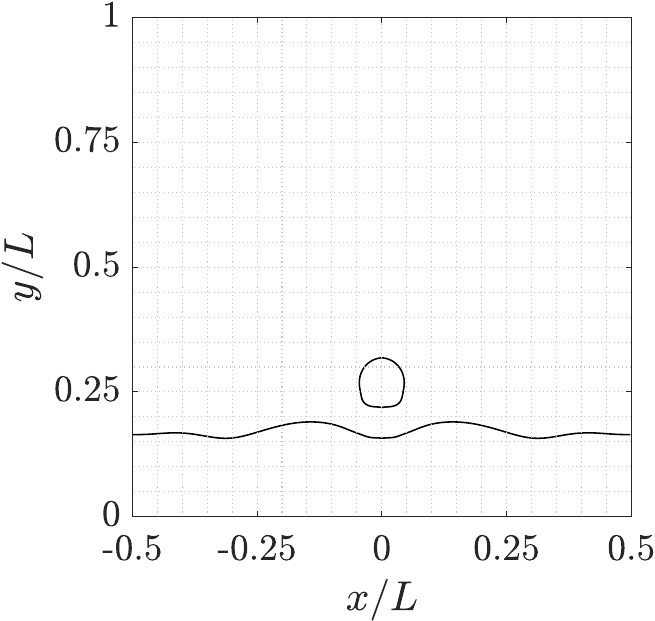}
\caption*{$t/T\sbs{o} = 1.9$}
\end{subfigure}
\begin{subfigure}{0.32\textwidth}
\includegraphics[trim = 0 0 0 0, clip, width =47mm]{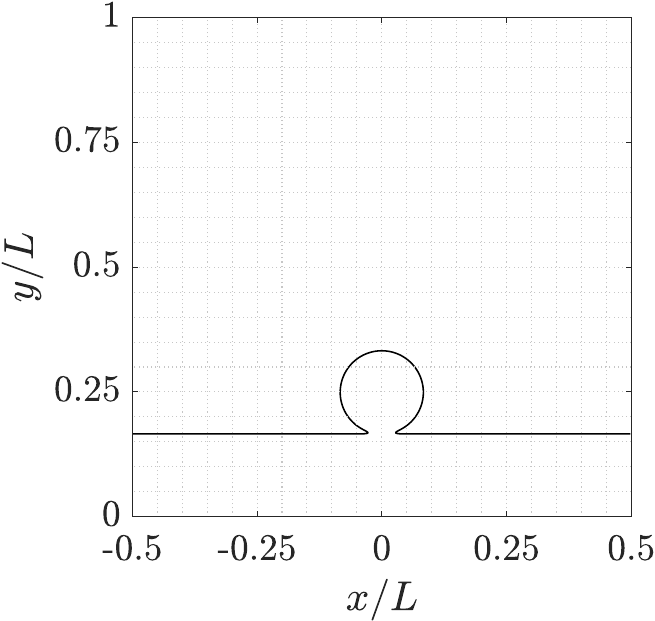}
\caption{\quad \quad $t/T\sbs{o} = 0$}
\label{initial_condition2}
\end{subfigure}
\begin{subfigure}{0.32\textwidth}
\includegraphics[trim = 0 0 0 0, clip, width =47mm]{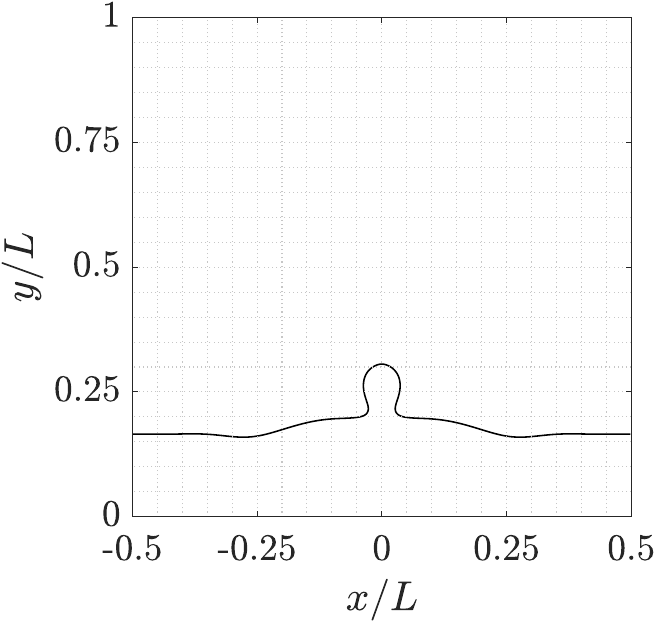}
\caption*{$t/T\sbs{o} = 1.65$}
\end{subfigure}
\begin{subfigure}{0.32\textwidth}
\includegraphics[trim = 0 0 0 0, clip, width =47mm]{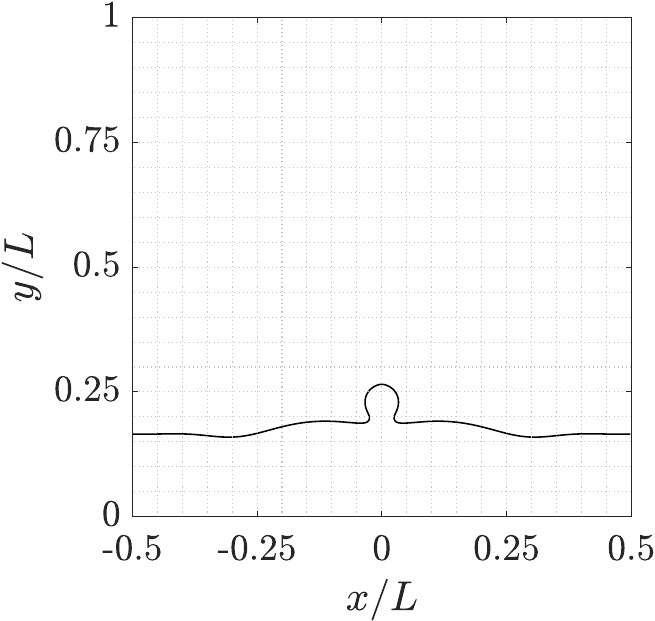}
\caption*{$t/T\sbs{o} = 1.9$}
\end{subfigure}
\caption{A thick black line illustrates the liquid interface in a two-dimensional view, depicting the interfacial evolution for coalescence/pinch-off between a liquid drop and the interface at various time intervals for two non-heated cases for the same Bond number $\mbox{Bo} = 0.01$, but with two different Ohnesorge numbers $(a)$ $\mbox{Oh} = 0.01$ and $(b)$ $\mbox{Oh} = 0.04$. The other parameters used for this simulation are $\tilde{\rho} = 1000$, and $\tilde{\mu} = 100$.}
\label{}
\end{figure}

For concreteness, we set the Bond number $\mbox{Bo} = 0.01$ and then study the effect of varying the Ohnesorge number by considering  $\mbox{Oh}= 0.01 \text{and}$ $0.04$. At the initial time $t/T\sbs{o} = 0$, where $T_o$ is given by $T_o = R/U_o = R/\sqrt{\sigma_o/(\rho_a R)}$, the problem setup for the non-heated case is depicted in the first subfigure on the left side of Figs.~\ref{initial_condition1} and~\ref{initial_condition2}. The simulation results for the subsequent stages of the coalescence process are illustrated in the second and third subfigures of Fig.~\ref{initial_condition1} at $\mbox{Oh} = 0.01$. Specifically, one snapshot captures the partial coalescence mechanism just before the actual coalescence at $t/T\sbs{o} = 1.65$, while another depicts the moment immediately after pinch-off at $t/T\sbs{o} = 1.9$. Furthermore, the second and third subfigures of Figure~\ref{initial_condition2} are results obtained with $\mbox{Oh}= 0.04$. They present variations in the interfacial motions during the coalescence process at the same time instances $t/T\sbs{o} = 1.65$ and $t/T\sbs{o} = 1.9$. Clearly, the case with the lower $\mbox{Oh}$ $(=0.01)$ undergoes pinch-off, while the higher $\mbox{Oh}$ $(=0.04)$ remains coalesced, illustrating significant differences in their outcomes. In particular, the greater the viscous forces relative to the surface tension force, the higher is the tendency of the drop to remain coalesced with the liquid layer; by contrast, as the former is weakened, the pinch-off outcome is favored since the surface tension force promotes the break-up as the neck becomes narrow enough.
\begin{figure}[H]
\centering
\begin{subfigure}{0.45\textwidth}
\includegraphics[trim = 90 65 80 100, clip, width =75mm]{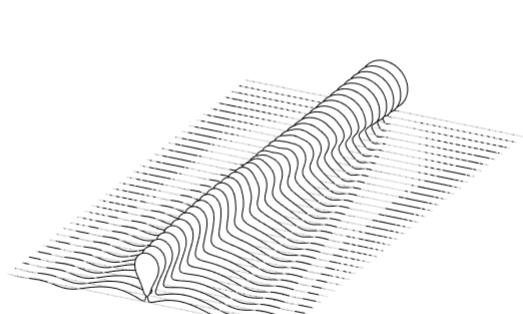}
\caption{$\mbox{Oh} = 0.01$}
\label{NH_Oh_01}
\end{subfigure}
\begin{subfigure}{0.45\textwidth}
\includegraphics[trim = 90 65 80 100, clip, width =75mm]{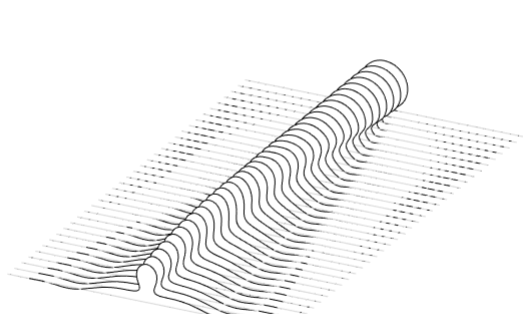}
\caption{$\mbox{Oh} = 0.04$}
\label{NH_Oh_04}
\end{subfigure}
\caption{Time evolution of a drop on an unheated liquid interface from the initial condition to pinch off event (back to front) contours of the order parameter, separated by $t/T\sbs{o} = 1/20$. The non-dimensional time duration for both cases, (a) $\mbox{Oh}= 0.01$ and (b) $\mbox{Oh}=0.04$, is $t/T\sbs{o} = 1.7$. The first pinch-off event for case $(a)$ is observed to occur at $t/T\sbs{o} = 1.7$, which is shown by the front-most contour in this subfigure. Here, the Bond number is $\mbox{Bo}=0.01$,  density ratio is $\tilde{\rho} = 1000$, and the viscosity ratio is $\tilde{\mu} = 100$.}
\label{NH_Oh_01_04}
\end{figure}
In Fig.~\ref{NH_Oh_01_04}, we show more comprehensive details involving a time sequence of drop dynamics (back to front) to compare two non-heated cases for the same Bond number, i.e., $\mbox{Bo} = 0.01$, and two different Ohnesorge numbers, i.e., $\mbox{Oh} = 0.01$ and $\mbox{Oh} = 0.04$, as shown in Figs.~\ref{NH_Oh_01} and~\ref{NH_Oh_04}, respectively. For small $Oh$, the pinch-off occurs as shown in Fig.~\ref{NH_Oh_01}. On the other hand, when we increase $\mbox{Oh}$, the pinch-off does not occur as shown in Fig.~\ref{NH_Oh_04}. We can thus conclude here that the effect of increasing the Ohnesorge number or increasing the viscous force relative to the surface tension force, can suppress the pinch-off mechanism.
\subsection{Normal fluid drop interacting with a heated fluid layer and its comparison with the non-heated case}
\begin{figure}[H]
\centering
\begin{subfigure}{0.465\linewidth}
\includegraphics[trim = 0 0 0 0,clip, width =62mm]{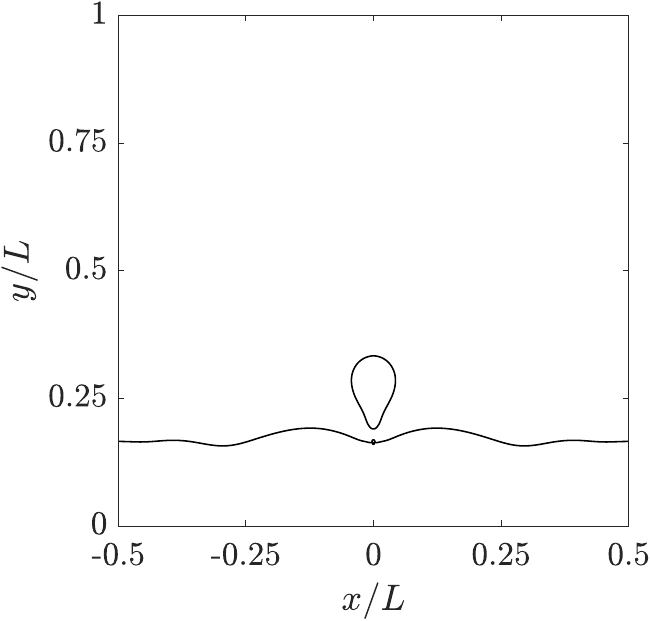}
\caption{Non-heated case}
\end{subfigure}
\begin{subfigure}{0.495\linewidth}
\includegraphics[trim = 0 0 0 20,clip, width =83mm]{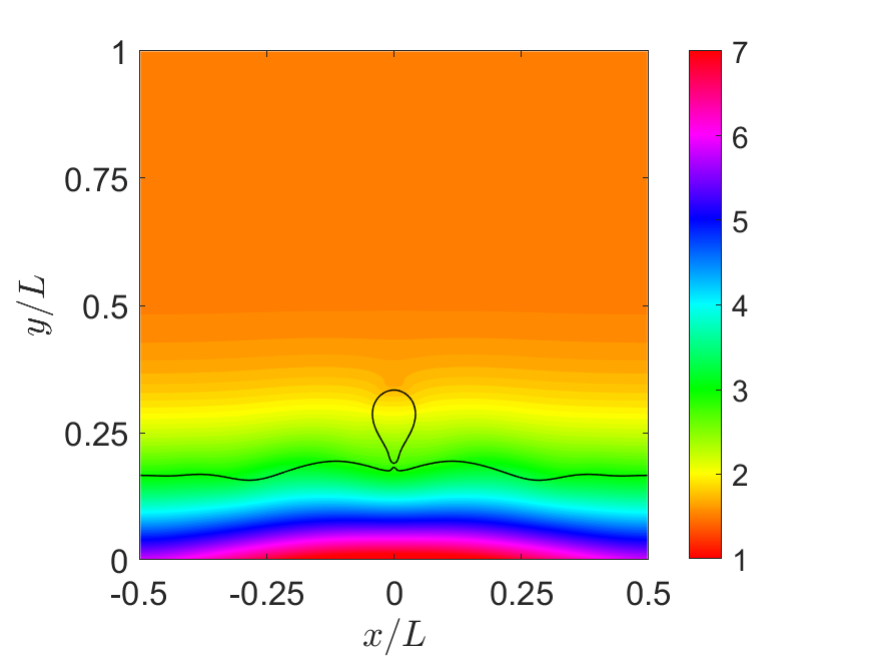}
\caption{NF heated case}
\end{subfigure}
\caption{Comparison of the contours of the order parameter presented at the onset of pinch-off of a NF drop from a liquid layer between (a) a non-heated case and (b) a heated case. For the heated case, $M_1 = 0.1$ and $Q = 0.02$. For both cases, the  non-dimensional time, $t/T\sbs{o} = 1.75$, and $M_2 = 0$, $\mbox{Bo} = 0.01$, $\mbox{Oh} = 0.01$, $\tilde{\rho} = 1000$, and $\tilde{\mu} = 100$. Color contours in the right figure represents the temperature field.}
\label{DOLI_NH_NF_T17_M1_01_q0_02_t_To_1_75}
\end{figure}

\begin{figure}[H]
\centering
\begin{subfigure}{0.465\linewidth}
\includegraphics[trim = 0 0 0 0,clip, width =75mm]{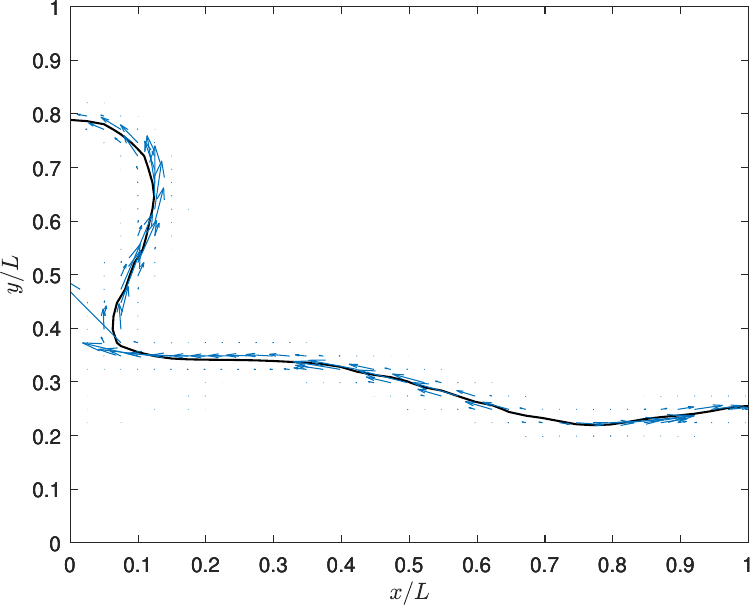}
\caption{Marangoni force}
\end{subfigure}
\begin{subfigure}{0.465\linewidth}
\includegraphics[trim = 0 0 0 0,clip, width =75mm]{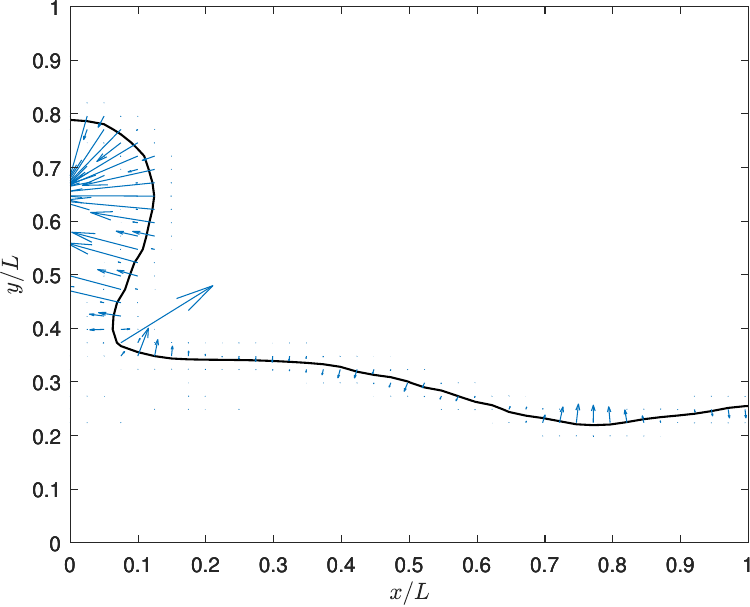}
\caption{Capillary force}
\end{subfigure}
\caption{Marangoni force $(a)$ and capillary force $(b)$ distribution around the interface for the NF case at the onset of the pinch-off. Here, $M_1 = 0.1$ and $Q = 0.02$. For both cases, the non-dimensional time, $t/T\sbs{o} = 1.5$, $M_2 = 0$, $\mbox{Bo} = 0.01$, $\mbox{Oh} = 0.01$, $\tilde{\rho} = 1000$, and $\tilde{\mu} = 100$.}
\label{Ma_Ca__Force_NF_T17_t_To_1_5}
\end{figure}

\begin{figure}[H]
\centering
\begin{subfigure}{0.465\linewidth}
\includegraphics[trim = 0 0 0 0,clip, width =75mm]{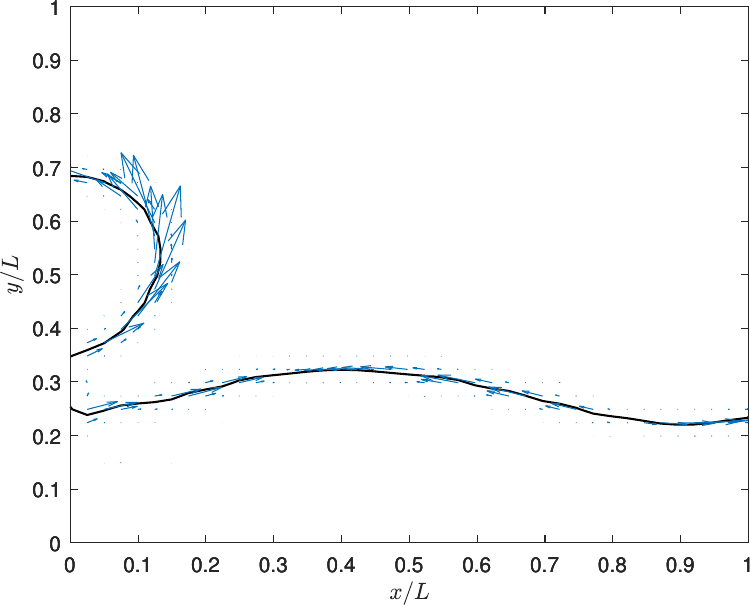}
\caption{Marangoni force}
\end{subfigure}
\begin{subfigure}{0.465\linewidth}
\includegraphics[trim = 0 0 0 0,clip, width =75mm]{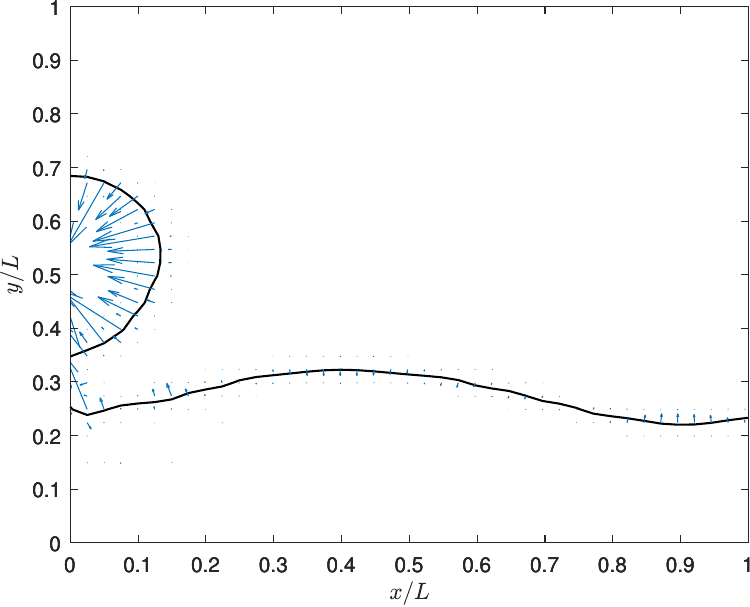}
\caption{Capillary force}
\end{subfigure}
\caption{Marangoni force $(a)$ and capillary force $(b)$ distribution around the interface for the NF case after the pinch-off. Here, $M_1 = 0.1$ and $Q = 0.02$. For both cases, the  non-dimensional time, $t/T\sbs{o} = 1.9$, $M_2 = 0$, $\mbox{Bo} = 0.01$, $\mbox{Oh} = 0.01$, $\tilde{\rho} = 1000$, and $\tilde{\mu} = 100$.}
\label{Ma_Ca__Force_NF_T17_t_To_1_9}
\end{figure}

Here, we show the interaction and pinch-off of a NF drop from a NF heated layer as compared with the corresponding non-heated case. In this regard, we require the linear coefficient of the surface tension variation with the temperature to be non-zero, i.e., $\sigma_{T} \neq 0$ or $M_1 \neq 0$, while the quadratic coefficient of the surface tension is zero, i.e., $\sigma_{TT} = 0$ or $M_2 = 0$. In the simulation, we set $M_1 = 0.1$ (and $M_2=0$) and the dimensionless heat flux $Q = 0.02$ for the heated case where $L_q/R=12$, and $\mbox{Bo} = 0.01$, $\mbox{Oh} = 0.01$, $\tilde{\rho} = 1000$, and $\tilde{\mu} = 100$ for both cases. Figure~\ref{DOLI_NH_NF_T17_M1_01_q0_02_t_To_1_75} shows a comparison mentioned above for the non-dimensional time of $t/T\sbs{o} = 1.75$. It can be seen that for the heated NF case, the pinch-off happens later when compared to the non-heated NF case. This result, the thermal delay of drop coalescence, is already claimed in~\cite{geri2017thermal}. This delay in the pinch-off of the drop for NFs happens because the Marangoni stress at the interface is directed toward the cold region, delaying the pinch-off of the drop. To illustrate this tangential stress modulated process for the heated case, Figs.~\ref{Ma_Ca__Force_NF_T17_t_To_1_5} and~\ref{Ma_Ca__Force_NF_T17_t_To_1_9} show the Marangoni (tangential) and capillary (normal) forces around the interface at the onset and following the pinch-off process, respectively. More explicitly, according to Eq.~(\ref{eqn6}), these forces can be identified as
\begin{equation*}
\qquad \bm{F}_{s} = \underbrace{  \sigma \kappa \bm{n} \delta_{s}}_{\text{Capillary force}} + \underbrace{ \bm{\nabla}_{s} \sigma \delta_{s}, }_{\text{Marangoni force}}
\end{equation*}
and since the surface tension varies with temperature as $\sigma = \sigma(T)$, their local variations in component form are given as the first and second terms in the right hand sides of Eq.~(\ref{eq:surfacetensionforcecomponents}). Clearly, in the NF heated case, the Marangoni force is always directed from the hot side to the cold side which decelerates the pinch-off process is absent in the non-heated case, thereby explaining the observed behavior.

\subsection{Self-rewetting fluid drop interacting with a heated fluid layer and its comparison with the non-heated case}
\begin{figure}[H]
\centering
\begin{subfigure}{0.45\linewidth}
\includegraphics[trim = 0 0 0 0,clip, width =62mm]{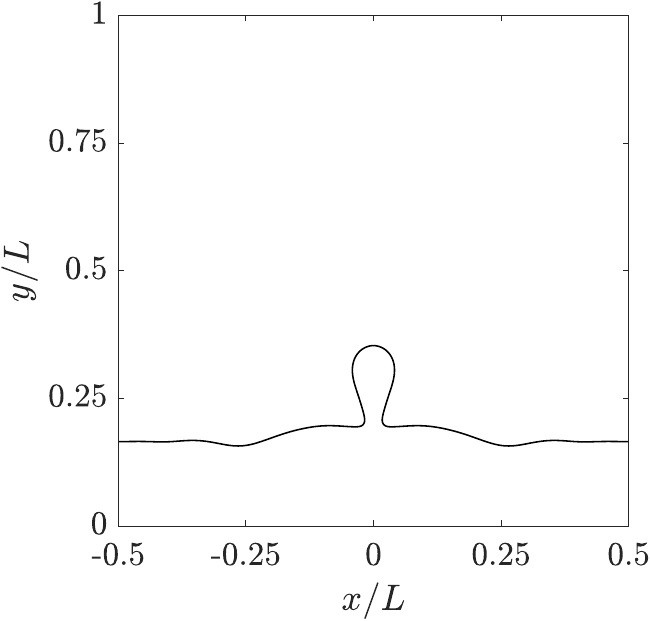}
\caption{Non-heated case}
\label{DOLI_NH_T17_t_To_1_5}
\end{subfigure}
\begin{subfigure}{0.45\linewidth}
\includegraphics[trim = 0 0 0 20,clip, width =83mm]{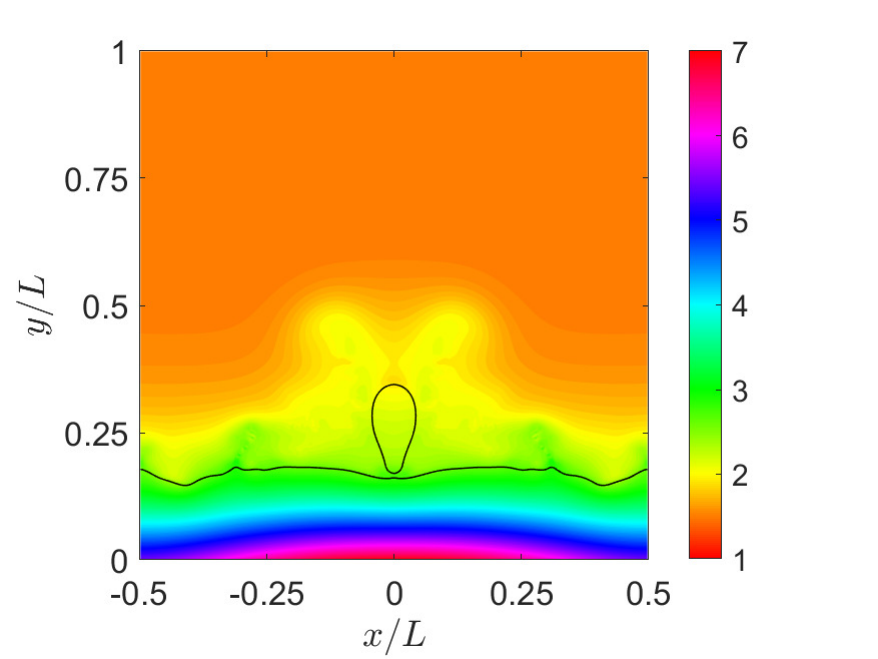}
\caption{SRF heated case}
\label{DOLI_SRF_T17_M2_5_q0_02_t_To_1_5}
\end{subfigure}
\caption{Comparison of the contours of the order parameter presented at the onset of pinch-off of a SRF drop from a liquid layer between (a) a non-heated case and (b) a heated case. For the SRFs heated case, $M_2 = 5$ and $Q = 0.02$. For both cases, the  non-dimensional time, $t/T\sbs{o} = 1.5$ and $M_1 = 0$, $\mbox{Bo} = 0.01$, $\mbox{Oh} = 0.01$, $\tilde{\rho} = 1000$, and $\tilde{\mu} = 100$. Color contours in the right figure represents the temperature field.}
\label{DOLI_NH_SRF_T17_M2_5_q0_02_t_To_1_5}
\end{figure}

\begin{figure}[H]
\centering
\begin{subfigure}{0.465\linewidth}
\includegraphics[trim = 0 0 0 0,clip, width =75mm]{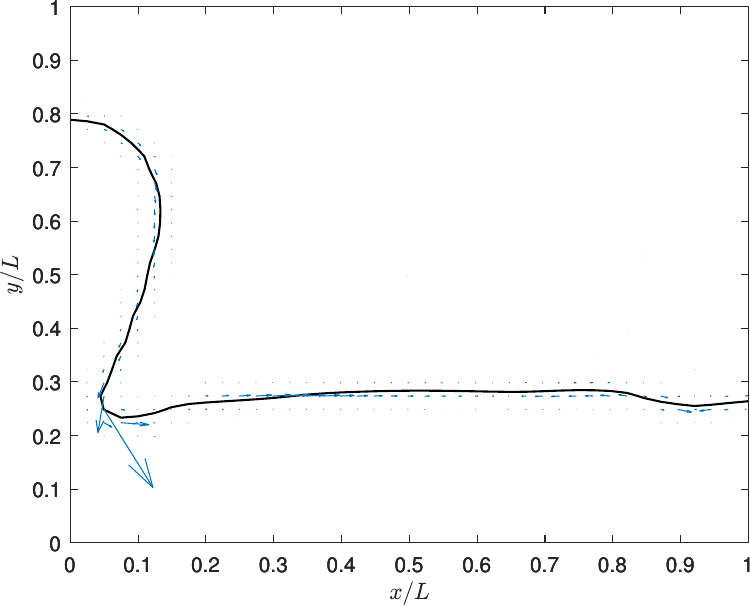}
\caption{Marangoni force}
\end{subfigure}
\begin{subfigure}{0.465\linewidth}
\includegraphics[trim = 0 0 0 0,clip, width =75mm]{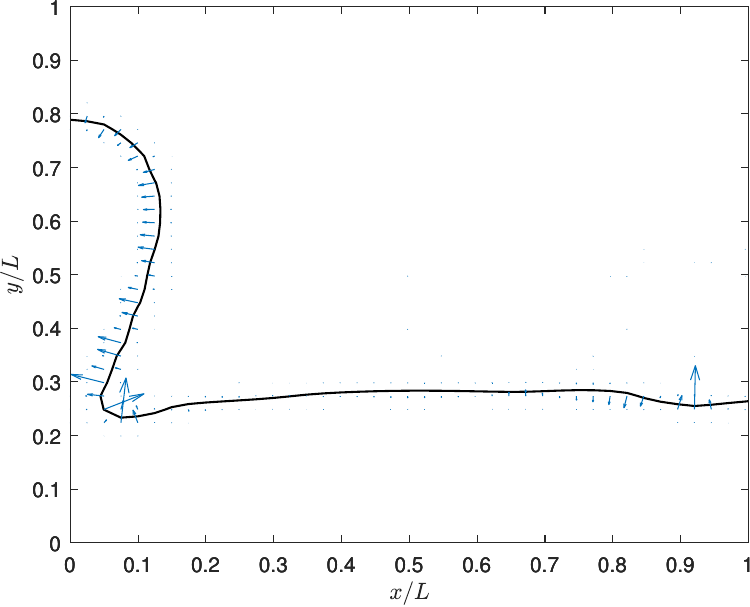}
\caption{Capillary force}
\end{subfigure}
\caption{Marangoni force $(a)$ and capillary force $(b)$ distribution around the interface for the SRF case at the onset of the pinch-off. Here, $M_2 = 5$ and $Q = 0.02$. For both cases, the non-dimensional time, $t/T\sbs{o} = 1.4$, $M_1 = 0$, $\mbox{Bo} = 0.01$, $\mbox{Oh} = 0.01$, $\tilde{\rho} = 1000$, and $\tilde{\mu} = 100$.}
\label{Ma_Ca__Force_SRF_T17_t_To_1_4}
\end{figure}

\begin{figure}[H]
\centering
\begin{subfigure}{0.465\linewidth}
\includegraphics[trim = 0 0 0 0,clip, width =75mm]{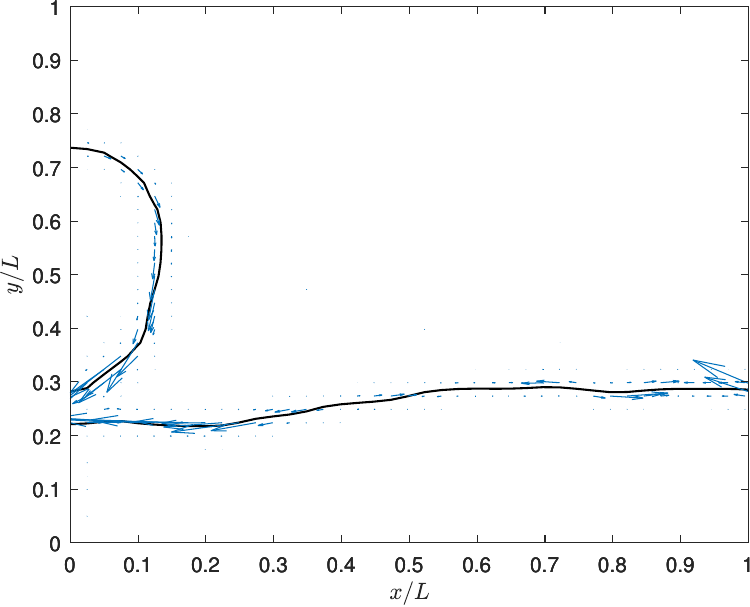}
\caption{Marangoni force}
\end{subfigure}
\begin{subfigure}{0.465\linewidth}
\includegraphics[trim = 0 0 0 0,clip, width =75mm]{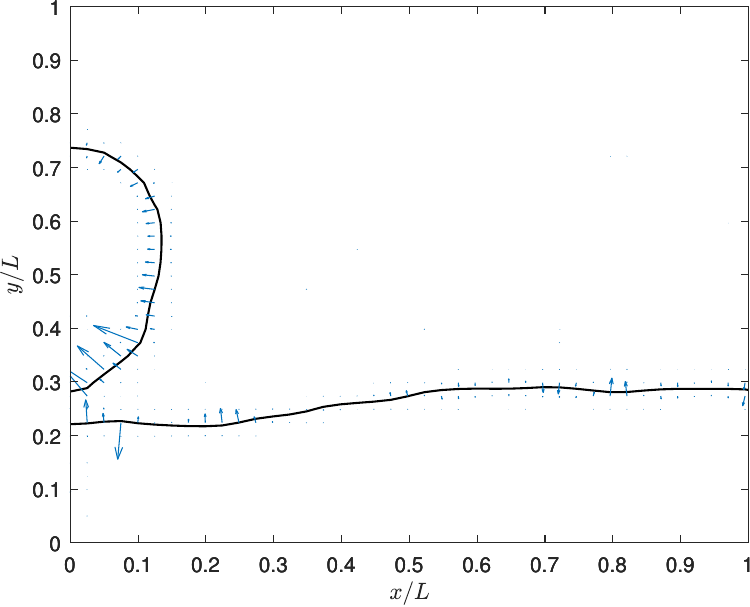}
\caption{Capillary force}
\end{subfigure}
\caption{Marangoni force $(a)$ and capillary force $(b)$ distribution around the interface for the SRF case after the pinch-off. Here, $M_2 = 5$ and $Q = 0.02$. For both cases, the non-dimensional time, $t/T\sbs{o} = 1.6$, $M_1 = 0$, $\mbox{Bo} = 0.01$, $\mbox{Oh} = 0.01$, $\tilde{\rho} = 1000$, and $\tilde{\mu} = 100$.}
\label{Ma_Ca__Force_SRF_T17_t_To_1_6}
\end{figure}

Next, let us compare the interaction and pinch-off of a SRF drop from a SRF heated layer as compared with the corresponding non-heated case. Here, we consider $M_2 = 5$ (and $M_1=0$) and the dimensionless heat flux $Q = 0.02$ for the heated case where $L_q/R=12$. Moreover, the choices of the other properties for both cases are the same as before, i.e., $\mbox{Bo} = 0.01$, $\mbox{Oh} = 0.01$, $\tilde{\rho} = 1000$, and $\tilde{\mu} = 100$. Figure~\ref{DOLI_NH_SRF_T17_M2_5_q0_02_t_To_1_5} shows snapshots of the drop liquid layer interaction for the heated and non-heated cases at a non-dimensional time of $t/T\sbs{o} = 1.5$, which is the earlier than the time instant shown in Fig.~\ref{DOLI_NH_NF_T17_M1_01_q0_02_t_To_1_75}. Clearly, while it is still in the process of pinching off for the non-heated case, the drop has already pinched off for the case of heated SRF drop, which indicates that the pinch-off occurs sooner in the SRFs case when compared to the NF case (see Fig.~\ref{DOLI_NH_NF_T17_M1_01_q0_02_t_To_1_75}) under nonuniform interfacial heating. Figures~\ref{Ma_Ca__Force_SRF_T17_t_To_1_4} and~\ref{Ma_Ca__Force_SRF_T17_t_To_1_6} show the Marangoni force distribution (along with the capillary force distribution) around the interface for the SRF case at the onset of the pinch-off and after the pinch-off, respectively. For the heated SRFs case, the Marangoni stress (or the resulting flow) on the interface is directed toward the hot region, accelerating the process of thinning the neck region and speeding up the pinch-off of the drop.

\begin{figure}[H]
\centering
\begin{subfigure}{0.45\textwidth}
\includegraphics[trim =0 0 0 0, clip, width =70mm]{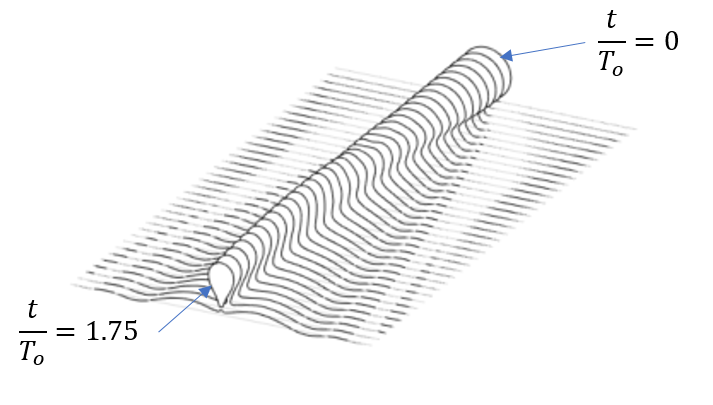}
\caption{NF ($M_1 = 0.1$)}
\end{subfigure}
\begin{subfigure}{0.45\textwidth}
\includegraphics[trim =  0 0 0 0, clip, width =70mm]{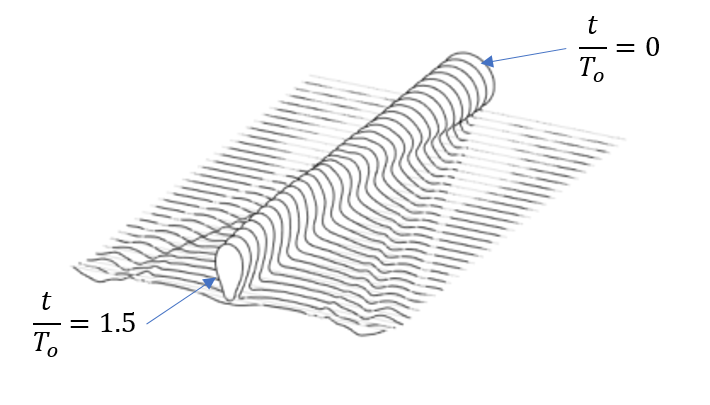}
\caption{SRF ($M_2 = 5$)}
\end{subfigure}
\caption{Time evolution of a drop on a heated liquid interface from the initial condition to pinch off event (back to front) contours of the order parameter, separated by $t/T\sbs{o} = 1/20$. $(a)$ NF with $M_1 = 0.1$, the non-dimensional time duration for NF is $t/T\sbs{o} = 1.75$.  $(b)$ SRF with $M_2 = 5$, the non-dimensional time duration for SRF is $t/T\sbs{o} = 1.5$. For both cases, $Q = 0.02$, $Bo=0.01$, $\mbox{Oh} = 0.01$, $\tilde{\rho} = 1000$, and $\tilde{\mu} = 100$.}
\label{fig:timesequenceeventsNF_SRF}
\end{figure}

\begin{figure}[H]
\centering
\begin{subfigure}{0.325\textwidth}
\includegraphics[trim = 0 0 0 0, clip, width =45mm]{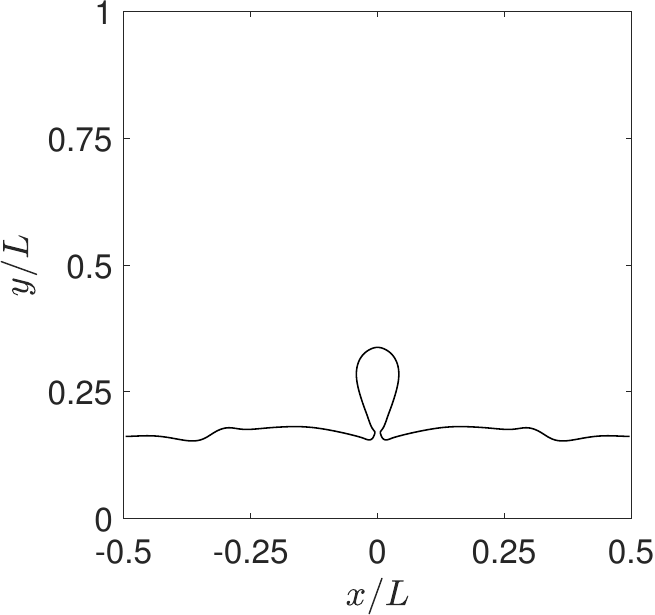}
\caption{$6R \times 6R$}
\end{subfigure}
\begin{subfigure}{0.325\textwidth}
\includegraphics[trim = 0 0 0 0, clip, width =45mm]{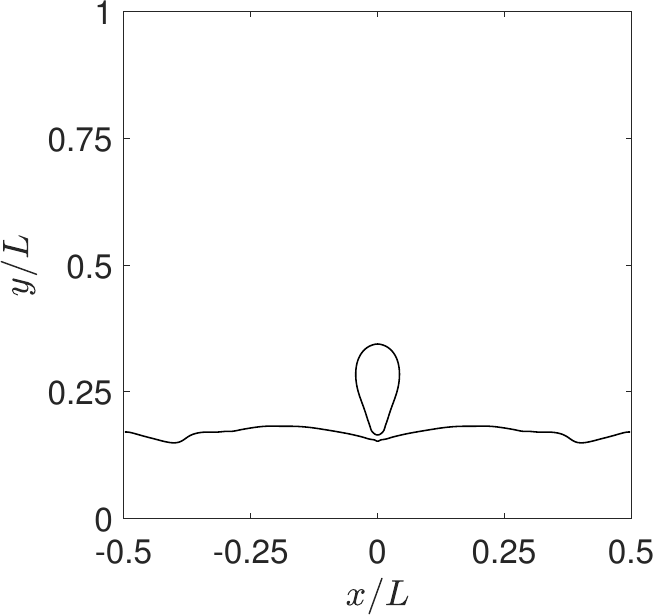}
\caption{$12R \times 12R$}
\end{subfigure}
\begin{subfigure}{0.325\textwidth}
\includegraphics[trim = 0 0 0 0, clip, width =45mm]{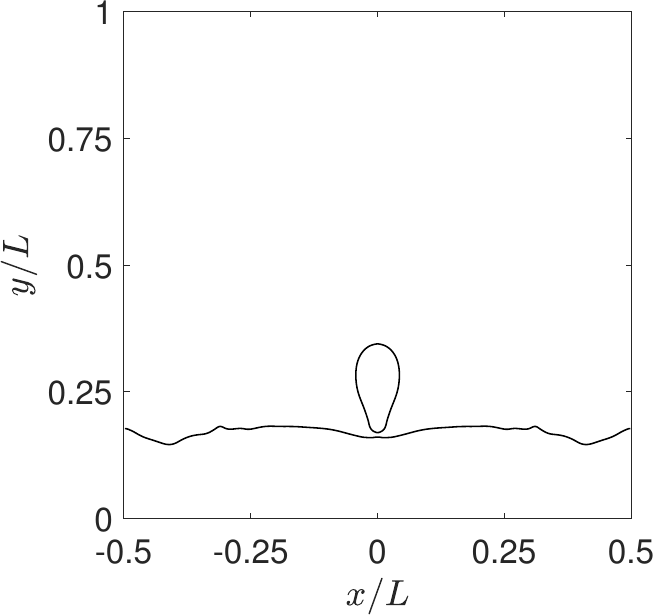}
\caption{$14R \times 14R$}
\end{subfigure}
\caption{Grid independence test results for the SRF case at the  non-dimensional time $t/T\sbs{o} = 1.5$ with $M_2 = 5$, $Q = 0.02$, $M_1 = 0$, $\mbox{Bo} = 0.01$, $\mbox{Oh} = 0.01$, $\tilde{\rho} = 1000$, and $\tilde{\mu} = 100$ for lower grid resolution ($6R \times 6R$), chosen grid resolution ($12R \times 12R$) and higher grid resolution ($14R \times 14R$), where $R=50$.}
\label{fig:gridindependencetest}
\end{figure}

Also, the earlier break up of the ligament from the SRF liquid layer seen in Fig.~\ref{DOLI_SRF_T17_M2_5_q0_02_t_To_1_5} is facilitated by the formation and evolution of capillary waves and the generation of more intense pressure differences across the neck region. This observation is further clarified as follows. Figure~\ref{fig:timesequenceeventsNF_SRF} presents the interface contours over time for both the NF case with $M_1=0.1$ as a reference and the SRF case with $M_2 = 5$ in the non-dimensional time intervals of $t/T\sbs{o} = 1/20$. The time sequence of events reveals that the formation of capillary waves assist in the break up process for SRF which occurs at $t/T\sbs{o}=1.5$; on the other hand, with the NF, due to the differences in the Marangoni force when compared to the former as noted above, the onset of such fragmentation does not occur until at a later time (i.e., $t/T\sbs{o}=1.75$). Moreover, in order to ensure that our results are independent of the choice of the grid resolution, in addition to the chosen resolution of $12R \times 12R$, we performed an additional set of simulations for the SRF case with lower resolution of $6R \times 6R$ and a higher resolution of $14R \times 14R$, where $R=50$. The results of these simulations are depicted in Fig.~\ref{fig:gridindependencetest} for the time instant $t/T\sbs{o} = 1.5$. Clearly, while the lower resolution case shows that the drop has not pinched off indicating the simulation is underresolved, the simulation results for the chosen resolution ($12R \times 12R$) and the higher resolution ($14R \times 14R$) cases agree with one another in that both result in the onset of pinch off. That is, a further increase in the resolution beyond the chosen grid resolution does not modify the results, which confirms the grid independence of the simulation results.

\begin{figure}[H]
\centering
\begin{subfigure}{0.495\linewidth}
\includegraphics[trim = 0 0 0 0,clip, width =75mm]{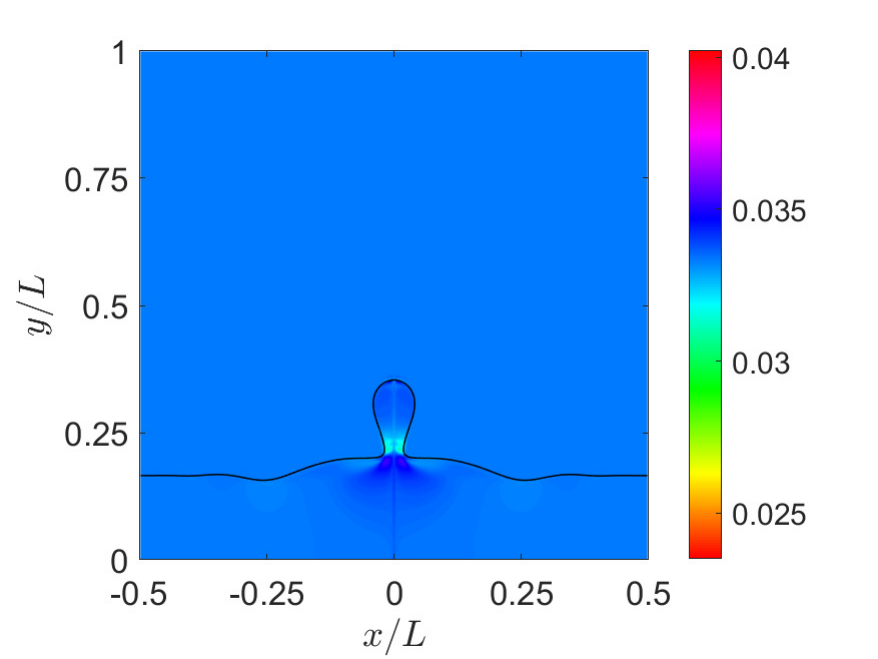}
\caption{NF before the pinch-off, $t/T\sbs{o} = 1.5$}
\end{subfigure}
\begin{subfigure}{0.495\linewidth}
\centering
\includegraphics[trim = 0 0 0 0, clip, width =75mm]{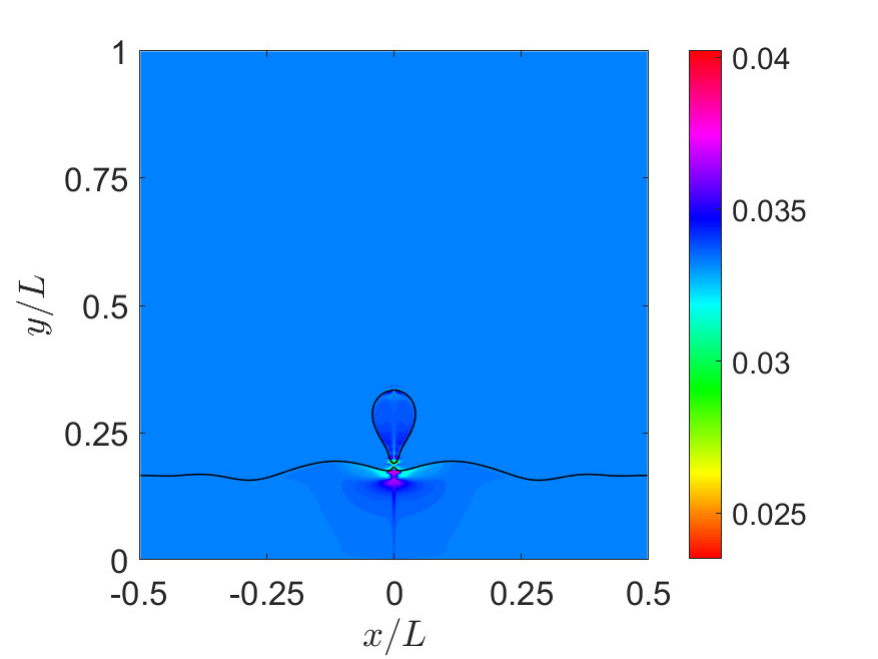}
\caption{NF at the pinch-off, $t/T\sbs{o} = 1.75$}
\end{subfigure}
\begin{subfigure}{0.495\linewidth}
\includegraphics[trim = 0 0 0 0,clip, width =75mm]{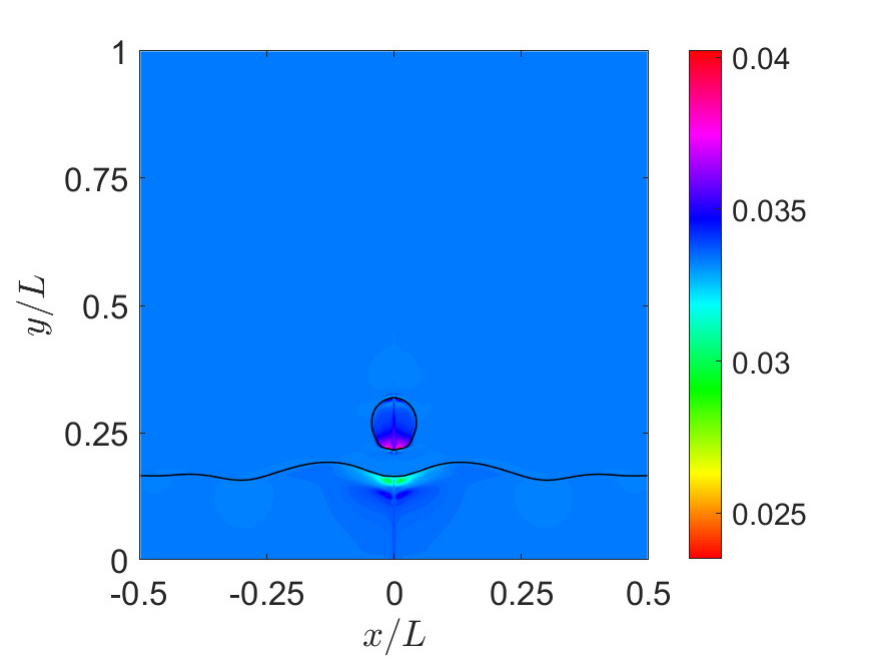}
\caption{NF after the pinch-off, $t/T\sbs{o} = 1.9$}
\end{subfigure}
\caption{Pressure contours for the NF case with $M_1 = 0.1$ and $Q = 0.02$. In addition, $\mbox{Bo} = 0.01$, $\mbox{Oh} = 0.01$, $\tilde{\rho} = 1000$, and $\tilde{\mu} = 100$.}
\label{fig:NFpressure}
\end{figure}

\begin{figure}[H]
\centering
\begin{subfigure}{0.495\linewidth}
\includegraphics[trim = 0 0 0 0,clip, width =75mm]{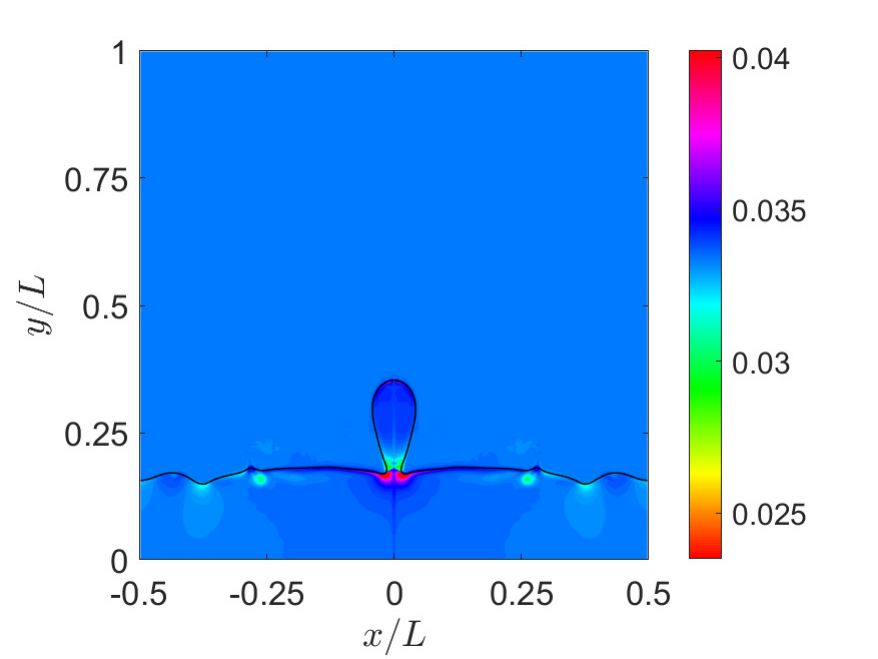}
\caption{SRF before the pinch-off, $t/T\sbs{o} = 1.4$}
\end{subfigure}
\begin{subfigure}{0.495\linewidth}
\centering
\includegraphics[trim = 0 0 0 0, clip, width =75mm]{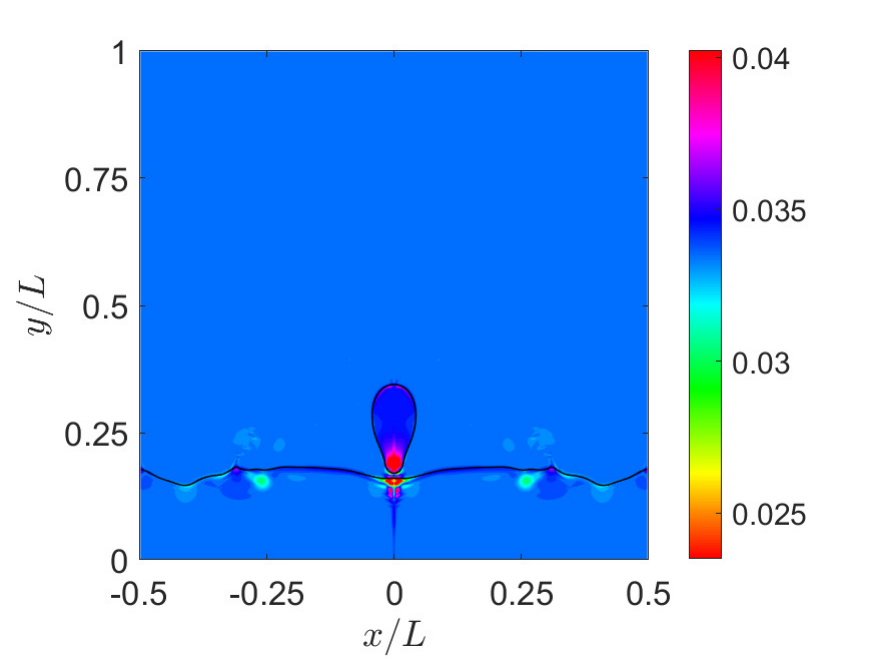}
\caption{SRF at the pinch-off, $t/T\sbs{o} = 1.5$}
\end{subfigure}
\begin{subfigure}{0.495\linewidth}
\includegraphics[trim = 0 0 0 0,clip, width =75mm]{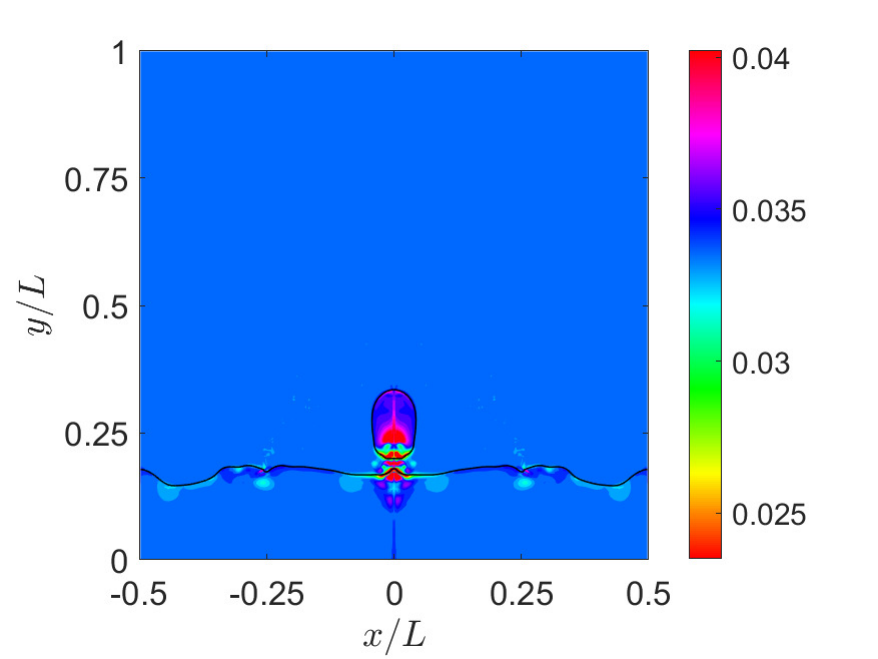}
\caption{SRF after the pinch-off, $t/T\sbs{o} = 1.6$}
\end{subfigure}
\caption{Pressure contours for the SRF case with $M_2 = 5$ and $Q = 0.02$. In addition, $\mbox{Bo} = 0.01$, $\mbox{Oh} = 0.01$, $\tilde{\rho} = 1000$, and $\tilde{\mu} = 100$.}
\label{fig:SRFpressure}
\end{figure}

Further insights can be gained by looking at and comparing the internal pressure distribution in the SRF case (see Fig~\ref{fig:SRFpressure}) with that in the reference NF case (see Fig~\ref{fig:NFpressure}). These two cases result in strikingly different pressure distributions which arise as a consequence of the differences in the surface tension behavior of these fluids. In particular, notice that the neck region in the SRF is associated with relatively higher pressure variations than those in the NF which promote the breaking up of the ligaments at an earlier instant in the former case.

In summary, it is evident that the pinch-off mechanism in SRFs is strikingly modulated in a different way than that in NFs. First, \emph{the pinch-off happens sooner} in SRFs, which is opposite to the case of NFs where \emph{the pinch-off happens at a somewhat later stage}. Second, the fluids on the interface seek to \emph{move towards the hotter region on the interface in the SRF case} while it moves towards the colder region in the NF case. These differences in the thermocapillarity between NFs and SRFs are a natural consequence of the opposite nature of action of the tangential Marangoni stress, which is produced when there is a positive (negative) surface tension gradient on the interface for SRFs (NFs).

\subsection{Effect of changing the various dimensionless characteristic parameters on the pinch-off/coalescence behavior}
\begin{figure}[H]
\centering
\includegraphics[trim = 0 0 0 0,clip, width = 75mm]{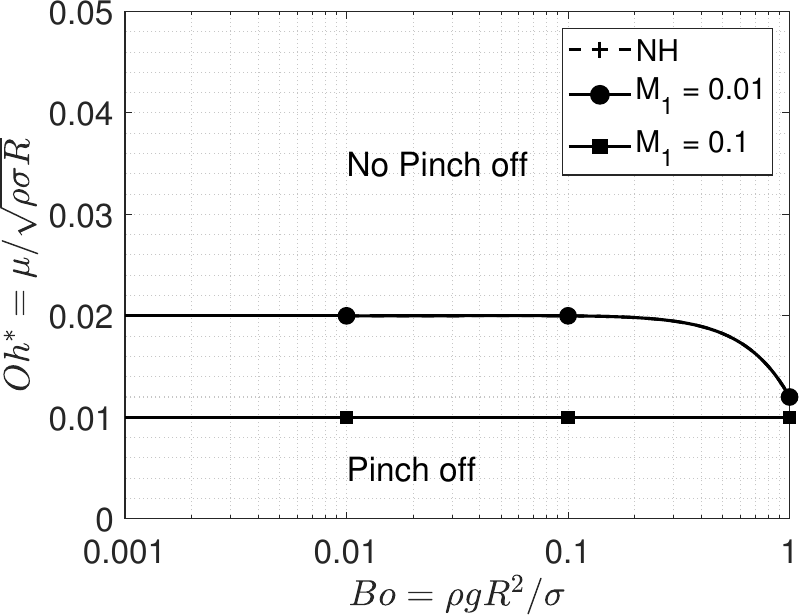}
\caption{The critical Ohnesorge number, $\mbox{Oh}^{\star}$, obtained from our axisymmetric central moment LBM multiphase flow simulations, as a function of the Bond number, $\mbox{Bo}$, for the non-heated (NH) case and heated case for normal fluid (NF) ($M_1 = 0.1$ and $M_1 = 0.01$). Here, $M_2 = 0$, $Q = 0.02$, $\tilde{\rho} = 1000$, and $\tilde{\mu} = 100$. Note that the red line for $M_1 = 0.01$ and the blue line for the non-heated (NH) case lie on top of each other.}
\label{Critical_Oh_NH_M1_01_M1_001_Q_02}
\end{figure}

Next, let us examine the effect of changing the dimensionless linear sensitivity coefficient of the surface tension variation with temperature, i.e., $M_1$ on the pinch-off/coalescence regime map, compared to the non-heated case. To do this, we use the following parameters: $Q = 0.02$, $L_q/R=12$, $\tilde{\rho} = 1000$, and $\tilde{\mu} = 100$ and vary $M_1$. For the case of NFs where only the linear coefficient is present (i.e., $M_2 = 0$), we use $M_1 = 0.01$ and $0.1$ which correspond for two different conditions/types of NFs. Figure~\ref{Critical_Oh_NH_M1_01_M1_001_Q_02} shows the critical Ohnesorge number, $\mbox{Oh}^{\star}$, curves, where above which, there is no pinch-off, and below pinch-off occurs as a function of the Bond number $\mbox{Bo}$ for the non-heated case (blue) and for the heated NF cases with $M_1=0.01$ (red) and $0.1$ (green). We found that a larger $M_1$ decreases the critical value of $\mbox{Oh}$, especially in the lower range of $\mbox{Bo}$ because the Marangoni force generated counteracts the pinch-off process. The large the $M_1$, the greater is this effect. In other words, increasing the linear surface tension sensitivity coefficient for normal fluids decreases the propensity of pinch-off as the corresponding region in the regime map is decreased at $M_1 = 0.1$ when compared to the other cases. These effects occur when $\mbox{Bo}<0.1$, while at higher $\mbox{Bo}$, when gravity effects significantly impede the dynamics associated surface tension forces, the role of variations in $M_1$ diminishes.

\begin{figure}[H]
\centering
\includegraphics[trim = 0 0 0 0,clip, width = 75mm]{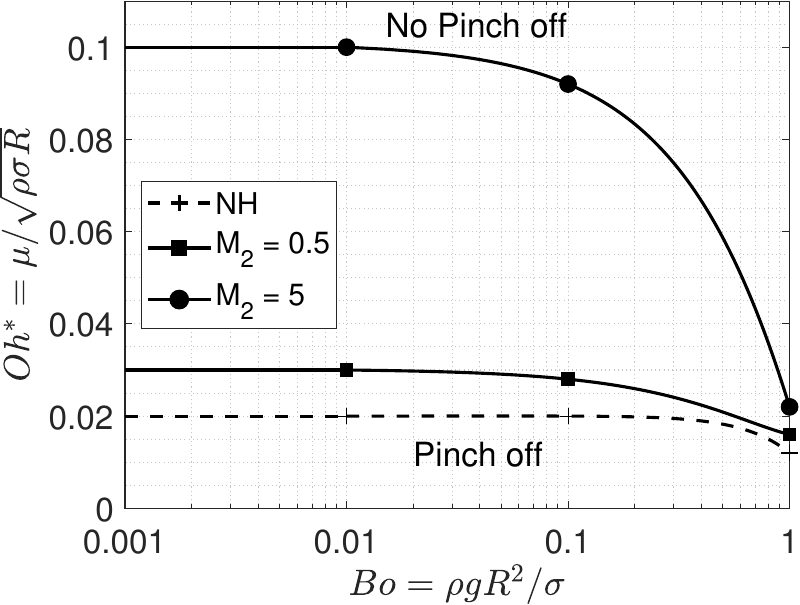}
\caption{The critical Ohnesorge number, $\mbox{Oh}^{\star}$, obtained from our axisymmetric central moment LBM multiphase flow simulations, as a function of the Bond number, $\mbox{Bo}$, for the non-heated (NH) case and heated case for self-rewetting fluid (SRF) ($M_2 = 0.5$ and $M_2 = 5$). Here, $M_1 = 0$, $Q = 0.02$, $\tilde{\rho} = 1000$, and $\tilde{\mu} = 100$.}
\label{Critical_Oh_NH_M2_05_M2_5_Q_02}
\end{figure}

Next, we will perform another study involving the effect of changing the dimensionless quadratic sensitivity coefficient of the surface tension variation with temperature, i.e., $M_2$ on the pinch-off/coalescence regime map, compared to the non-heated case. In this regard, we use two types of SRFs, by fixing $M_2 = 0.5$ and $M_2 = 5$. The other parameters we use in this study are $M_1 = 0$, $Q = 0.02$, $L_q/R=12$, $\tilde{\rho} = 1000$, and $\tilde{\mu} = 100$. Figure~\ref{Critical_Oh_NH_M2_05_M2_5_Q_02} shows the critical Ohnesorge number, $\mbox{Oh}^{\star}$, curves above which there is no pinch-off, and below that pinch-off occurs, as a function of the Bond number $\mbox{Bo}$ for the non-heated case (blue) and for the heated SRF cases with $M_2=0.5$ (green) and $5$ (red). Interestingly, for the higher value of the quadratic coefficient, i.e., $M_2 = 5$, the critical value of $\mbox{Oh}$ increases for the lower range of $\mbox{Bo}$ (when $\mbox{Bo}<0.1$) because greater Marangoni forces in SRFs is directed towards higher temperature zones in the interfacial region that promotes the pinch-off process. Thus, increasing the quadratic surface tension sensitivity coefficient for self-rewetting fluids from $M_2=0.5$ to $5.0$ dramatically increases the area in the $\mbox{Bo}-\mbox{Oh}$ regime map where pinch-off can occur; moreover, the non-heated case has the least region for the tendency to pinch-off. This behavior is opposite to that we observed in the previous section for the NF case. These contrasting effects between NFs and SRFs may be used to design strategies to control coalescence/pinch-off process via imposed nonuniform heating.

\begin{figure}[H]
\centering
\begin{subfigure}{0.45\linewidth}
\includegraphics[trim = 0 0 0 0,clip, width =65mm]{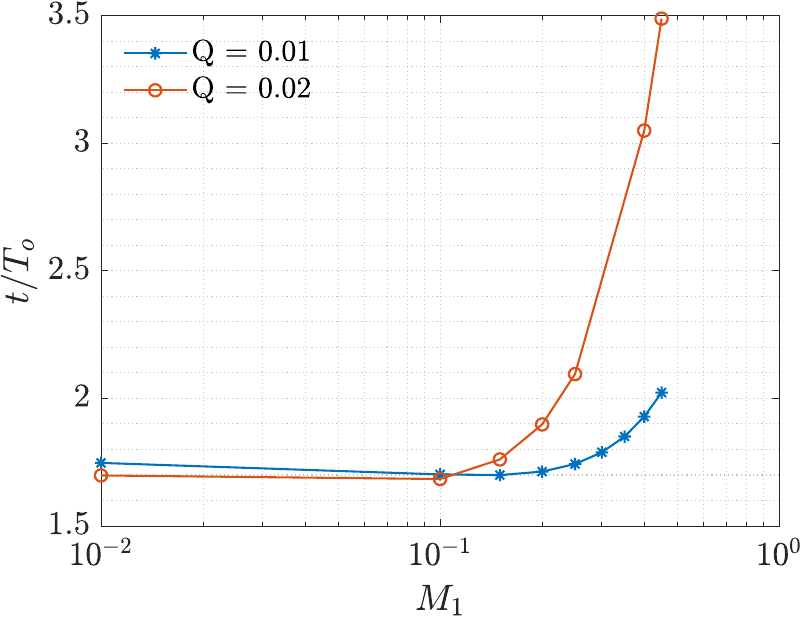}
\caption{Normal fluids (NFs)}
\end{subfigure}
\begin{subfigure}{0.45\linewidth}
\includegraphics[trim = 0 0 0 0,clip, width =65mm]{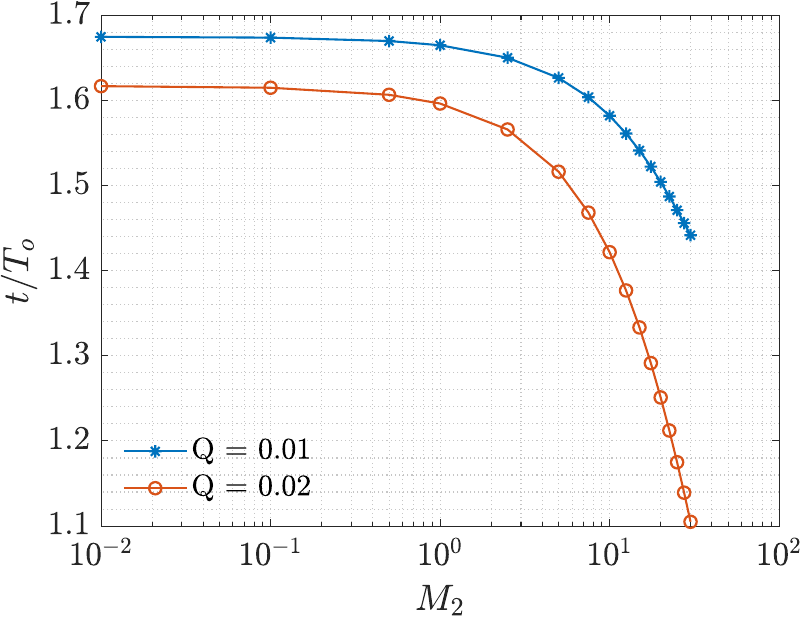}
\caption{Self-rewetting fluids (SRFs)}
\end{subfigure}
\caption{Dimensionless time duration $t/T_o$ for the onset of pinch-off for (a) normal fluids (NFs) with $M_2 = 0$ as a function of $M_1$ and (b) self-rewetting fluids (SRFs) with $M_1 = 0$ as a function of $M_2$ at $Q = 0.01$ and $Q = 0.02$. Here, $\mbox{Bo} = 0.1$, $\mbox{Oh} = 0.01$, $\tilde{\rho} = 1000$, and $\tilde{\mu} = 100$. The time scale $T_o$ is given by $T_o = R/U_o = R/\sqrt{\sigma_o/(\rho_a R)}$.}
\label{NF_SRF_pinch_off_time_Q_01_and_02}
\end{figure}

Finally, let us now study the effect of changing $M_1$ and $M_2$ for NF and SRF, respectively on the duration of the onset of pinch-off time for two different dimensionless heat fluxes, i.e., $Q = 0.01$, $Q = 0.02$ when $L_q/R=12$. In this regard, we plot the duration for the onset of pinch-off for NFs and SRFs as shown in Fig.~\ref{NF_SRF_pinch_off_time_Q_01_and_02}. Here, the rest of the parameters we use in this study are $\mbox{Bo} = 0.1$, $\mbox{Oh} = 0.01$, $\tilde{\rho} = 1000$, and $\tilde{\mu} = 100$. The choice of fluid is seen to have a significant impact on when the drop can start to pinch off. For NFs, the pinch-off time increases as the dimensionless surface tension coefficient $M_1$ increases. By contrast, the behavior is opposite to that for SRF, where the pinch-off time reduces as $M_2$ increases. Moreover, doubling the dimensionless heat flux $Q$ for the same $\mbox{Bo}$ and $\mbox{Oh}$, i.e., comparing $Q = 0.02$ to $Q = 0.01$, we notice that the drop pinches off sooner for the latter case, which is expected for SRFs and also pinches off for a wider range of parameters which is opposite in behavior to the case of NFs. This is due to the fact that increasing the heat flux $Q$ causes more pronounced interfacial heating with a concomitant increase in the tangential Marangoni stress that promotes the tendency to pinch-off in the SRF case or retards it in the case of NF case. In general, the effect $Q$ is greater for larger $M_1$ or $M_2$, as appropriate. In summary, there is a greater tendency to pinch-off for a broad range of conditions for SRFs when compared to NFs while also hastening the process in the former case.

\section{Results and discussion: Imposed temperature boundary condition case} \label{Temp BC}
\begin{figure}[H]
\centering
\includegraphics[trim = 0 0 0 0, clip, width =90mm]{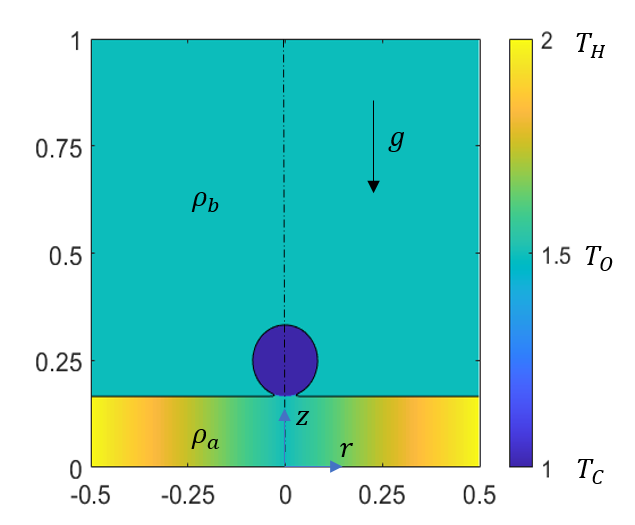}
\caption{Schematic of the initial condition of a drop on liquid interface for the heated case with an imposed non-uniform (linear) temperature distribution at the bottom boundary ($z=0$). Using an axisymmetric model, the axis of symmetry is the vertical centerline shown in the figure, and the entire domain is shown here for clarity.}
\label{Temp_heated_MODEL}
\end{figure}

In this section, we will illustrate how changing from the imposed heat flux studied in the previous section to a nonuniform surface temperature boundary condition on the bottom side of the domain influences the coalescence/pinch-off process during drop-liquid layer interaction. The setup of this problem is identical to that of Sec.~\ref{DOLI HF Section.1}, except for the heating conditions at the bottom boundary, where a nonuniform temperature distribution is imposed instead of heat flux. The heating conditions we used are shown in Fig.~\ref{Temp_heated_MODEL}, where the temperature on the surface increases linearly along the radial direction, is similar to the study performed by~\cite{geri2017thermal} that obtained an analytical solution for NFs in which the pool is heated from the sides maintained at a temperature $T_H$, the drop initially at a lower temperature $(T_C)$, and the reference temperature $T_{ref}$ is used everywhere else.

Throughout this study, unless explicitly stated otherwise, we maintain the following parameters at constant values: $T_H=2$ (hot temperature), $T_C=1$ (cold temperature), $T_{ref}=1.5$ (reference temperature), $R=50$ (drop radius), $\sigma_o = 0.005$ (surface tension coefficient at the reference temperature), $\tilde{\rho} = 1000$ (density ratio), $\tilde{\mu} = 100$ (dynamic viscosity ratio), and $\tilde{\alpha} = 1$ (thermal diffusivity ratio). Additionally, for the model parameters in the conservative Allen-Cahn equation used for interface tracking, we set the interface thickness $W$ to $5$ and the mobility coefficient $M_{\phi}$ to $0.1$.

\subsection{Normal fluid drop interacting with a heated fluid layer and its comparison with the non-heated case}
\begin{figure}[H]
\centering
\begin{subfigure}{0.465\linewidth}
\includegraphics[trim = 0 0 0 0,clip, width =62mm]{DOLI_NH_T17_t_To_1_75}
\caption{Non-heated case}
\end{subfigure}
\begin{subfigure}{0.495\linewidth}
\includegraphics[trim = 0 0 0 20,clip, width =83mm]{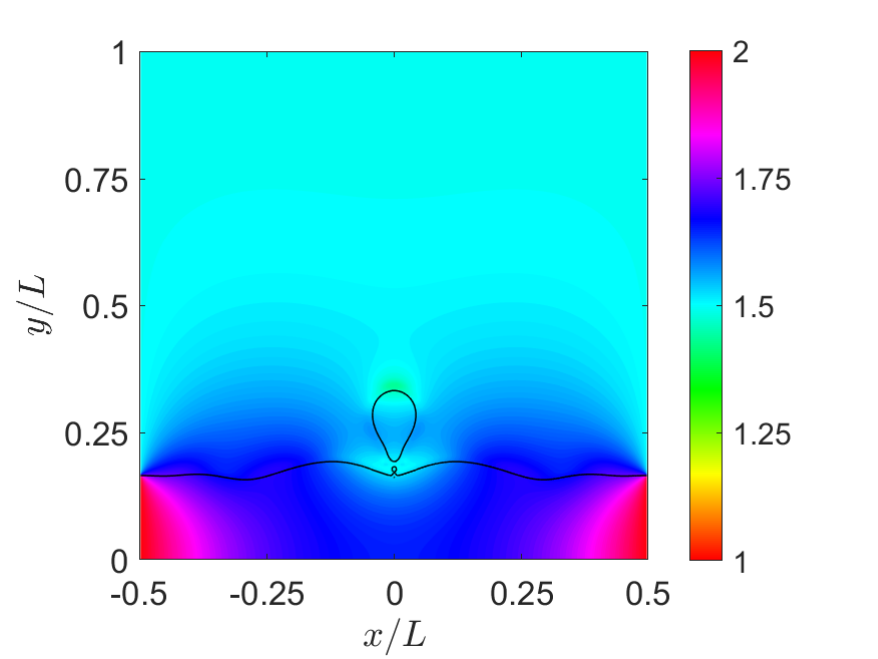}
\caption{NF heated case}
\end{subfigure}
\caption{Comparison of the contours of the order parameter presented at the onset of pinch-off of a NF drop from a liquid layer between (a) a non-heated case and (b) a heated case. For the heated case, $M_1 = 0.1$. For both cases, the non-dimensional time, $t/T\sbs{o} = 1.75$, and $M_2 = 0$, $\mbox{Bo} = 0.01$, $\mbox{Oh} = 0.01$, $\tilde{\rho} = 1000$, and $\tilde{\mu} = 100$. Color contours in the right figure represents the temperature field.}
\label{DOLI_NH_NF_T17_M1_01_t_To_1_75}
\end{figure}

We now present the interaction and pinch-off of a normal fluid (NF) drop from a heated NF layer and compare it with a scenario where no heating is applied. In our simulations, we set $M_1 = 0.1$ (and $M_2 = 0$) alongside with $\tilde{\rho} = 1000$ and $\tilde{\mu} = 100$ for both scenarios. The comparison at the non-dimensional time $t/T\sbs{o} = 1.75$ is depicted in Fig.~\ref{DOLI_NH_NF_T17_M1_01_t_To_1_75}, highlighting that the pinch-off in the heated NF scenario occurs later compared to the non-heated NF scenario, an effect previously documented by Geri~\emph{et al.}~\cite{geri2017thermal} as thermal delay in droplet coalescence. This postponement in droplet pinch-off for NFs is attributed to the Marangoni stress, which is oriented towards the cooler area, thereby delaying the droplet's pinch-off. This mechanism, influenced by the tangential force in the heated scenario, is similar to those illustrated previously in Figs.~\ref{Ma_Ca__Force_NF_T17_t_To_1_5} and~\ref{Ma_Ca__Force_NF_T17_t_To_1_9} before and after the pinch-off process, respectively. In the heated NF scenario, the Marangoni stress consistently moves from the hotter to the cooler side, a phenomenon not present in the non-heated scenario, thus explaining the observed behavior.

\subsection{Self-rewetting fluid drop interacting with a heated fluid layer and its comparison with the non-heated case}
\begin{figure}[H]
\centering
\begin{subfigure}{0.45\linewidth}
\includegraphics[trim = 0 0 0 0,clip, width =62mm]{DOLI_NH_T17_t_To_1_75}
\caption{Non-heated case}
\label{DOLI_NH_T17_t_To_1_75}
\end{subfigure}
\begin{subfigure}{0.45\linewidth}
\includegraphics[trim = 0 0 0 20,clip, width =83mm]{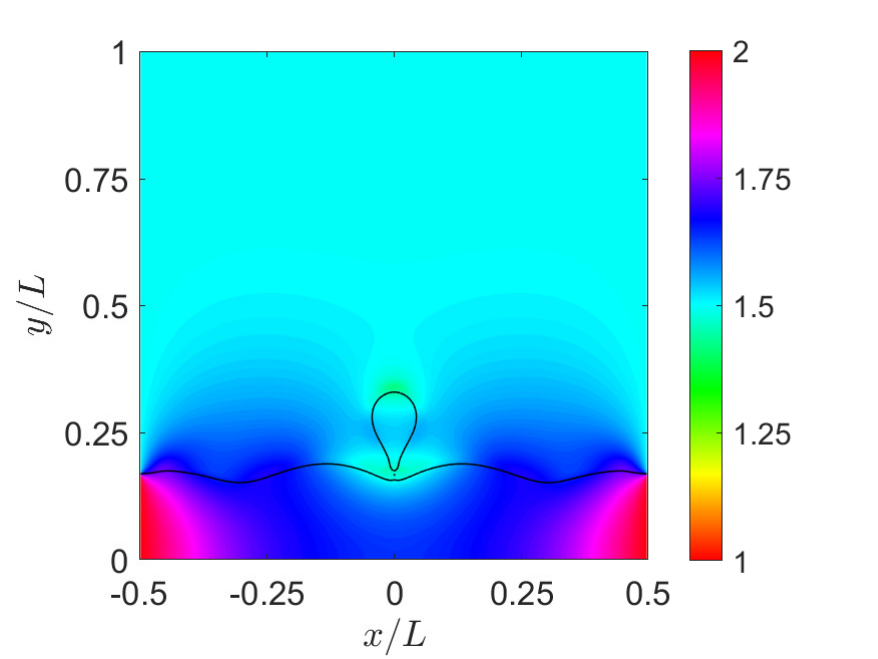}
\caption{SRF heated case}
\label{DOLI_SRF_T17_M2_5_t_To_1_75}
\end{subfigure}
\caption{Comparison of the contours of the order parameter presented at the onset of pinch-off of a SRF drop from a liquid layer between (a) a non-heated case and (b) a heated case. For the SRFs heated case, $M_2 = 5$. For both cases, the non-dimensional time, $t/T\sbs{o} = 1.75$, and $M_1 = 0$, $\mbox{Bo} = 0.01$, $\mbox{Oh} = 0.01$, $\tilde{\rho} = 1000$, and $\tilde{\mu} = 100$. Color contours in the right figure represents the temperature field.}
\label{DOLI_NH_SRF_T17_M2_5_t_To_1_75}
\end{figure}

Let us now examine the difference between the interaction and pinch-off of a SRF drop from a heated SRF layer and the comparable non-heated situation. For the heated scenario, we set $M_2 = 5$. Moreover, we specify the other parameters as $\mbox{Bo} = 0.01$, $\mbox{Oh} = 0.01$, $\tilde{\rho} = 1000$, and $\tilde{\mu} = 100$, just as the previous case. Figure~\ref{DOLI_NH_T17_t_To_1_75} and~\ref{DOLI_SRF_T17_M2_5_t_To_1_75}, respectively, present snapshots of the drop liquid layer interaction for the non-heated and heated situations at a non-dimensional time of $t/T\sbs{o} = 1.75$. As can be seen from Fig.~\ref{DOLI_NH_SRF_T17_M2_5_t_To_1_75}, the drop appears to be already pinched off for the heated SRF case, but it just gets pinched off for the non-heated case. This suggests that the pinch-off happens sooner in the SRFs case than in the NF case (see Fig.~\ref{DOLI_NH_NF_T17_M1_01_t_To_1_75}). The Marangoni force distribution (together with the capillary force distribution) around the interface for the SRF case at the start of the pinch-off and after the pinch-off are analogous to those shown in Figs.~\ref{Ma_Ca__Force_SRF_T17_t_To_1_4} and~\ref{Ma_Ca__Force_SRF_T17_t_To_1_6}, respectively. Thinning of the neck region and accelerating the pinch-off of the drop is the effect of the Marangoni stress (and the resultant flow) on the interface directed toward the hot region in the heated SRFs case.
Additionally, capillary waves occur and help speed up the ligament's quicker break up from the SRF liquid layer.


\subsection{Effect of changing the various dimensionless characteristic parameters on the pinch-off/coalescence behavior}
\begin{figure}[H]
\centering
\includegraphics[trim = 0 0 0 0,clip, width = 75mm]{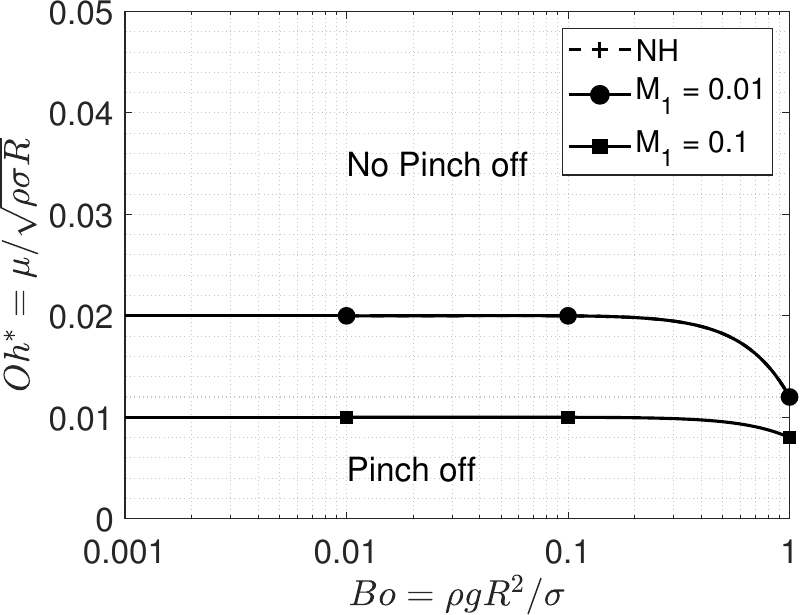}
\caption{The critical Ohnesorge number, $\mbox{Oh}^{\star}$ obtained from our axisymmetric central moment LBM multiphase flow simulations, as a function of the Bond number, $\mbox{Bo}$, for the non-heated (NH) case and heated case for normal fluid (NF) ($M_1 = 0.1$ and $M_1 = 0.01$). Here, $M_2 = 0$, $\tilde{\rho} = 1000$, and $\tilde{\mu} = 100$. Note that the red line for $M_1 = 0.01$ and the blue line for the non-heated (NH) case lie on top of each other.}
\label{Critical_Oh_NH_M1_01_M1_001}
\end{figure}

Next, we investigate the effect of varying the dimensionless linear sensitivity coefficient of the temperature-dependent surface tension variation, or $M_1$, on the pinch-off/coalescence regime map in comparison to the non-heated scenario. In order to do this, we set $\tilde{\rho} = 1000$ and $\tilde{\mu} = 100$ for each case. In the case of NFs with only a linear coefficient (i.e., $M_2 = 0$), we utilize $M_1 = 0.01$ and $0.1$, which correspond to two different conditions/types of NFs. The critical Ohnesorge number, $\mbox{Oh}^{\star}$, is displayed in Figure~\ref{Critical_Oh_NH_M1_01_M1_001}. Its behavior is found to be similar to that shown earlier in Fig.~\ref{Critical_Oh_NH_M1_01_M1_001_Q_02} for the imposed heat flux condition for nonuniform heating. Put simply, increasing the linear coefficient of surface tension's response to temperature in normal fluids reduces the likelihood of drop detachment. This is evidenced by a smaller region on the regime map for pinch-off at $M_1 = 0.1$ versus other scenarios.

\begin{figure}[H]
\centering
\includegraphics[trim = 0 0 0 0,clip, width = 75mm]{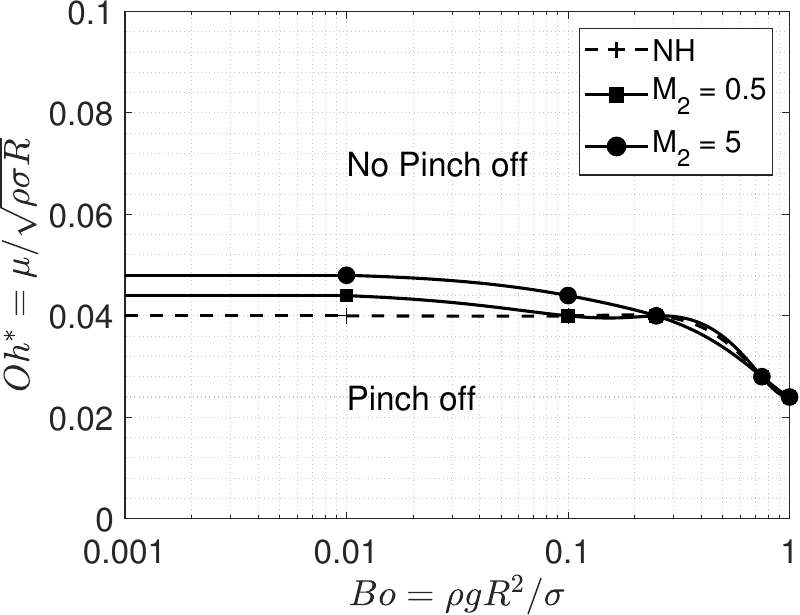}
\caption{The critical Ohnesorge number, $\mbox{Oh}^{\star}$ obtained from our axisymmetric central moment LBM multiphase flow simulations, as a function of the Bond number, $\mbox{Bo}$, for the non-heated (NH) case and heated case for self-rewetting fluid (SRF) ($M_2 = 0.5$ and $M_2 = 5$). Here, $M_1 = 0$, $\tilde{\rho} = 1000$, and $\tilde{\mu} = 100$.}
\label{Critical_Oh_NH_M2_05_M2_5}
\end{figure}

We now report the effect of changing the dimensionless quadratic sensitivity coefficient of the surface tension variation with temperature, or $M_2$, on the pinch-off/coalescence regime map in comparison to the non-heated case. In this regard, we use two different SRF types in this regard, fixing $M_2 = 0.5$ and $M_2 = 5$. In this study, we use $M_1 = 0$, $\tilde{\rho} = 1000$ and $\tilde{\mu} = 100$ as additional parameters. The critical Ohnesorge number, $\mbox{Oh}^{\star}$, is displayed in Fig.~\ref{Critical_Oh_NH_M2_05_M2_5} as a function of the Bond number $\mbox{Bo}$ for both the heated SRF cases with $M_2=0.5$ (green) and $5$ (red), as well as for the non-heated case (blue). Unlike the NF case as discussed in the previous section, the use of SRF broadens the conditions when the drop can pinch-off as $M_2$ is increased. However, these observations, while consistent, are not as dramatic as those reported for the imposed heat flux boundary condition (see Fig.~\ref{Critical_Oh_NH_M2_05_M2_5_Q_02}).

\begin{figure}[H]
\centering
\begin{subfigure}{0.45\linewidth}
\includegraphics[trim = 0 0 0 0,clip, width =65mm]{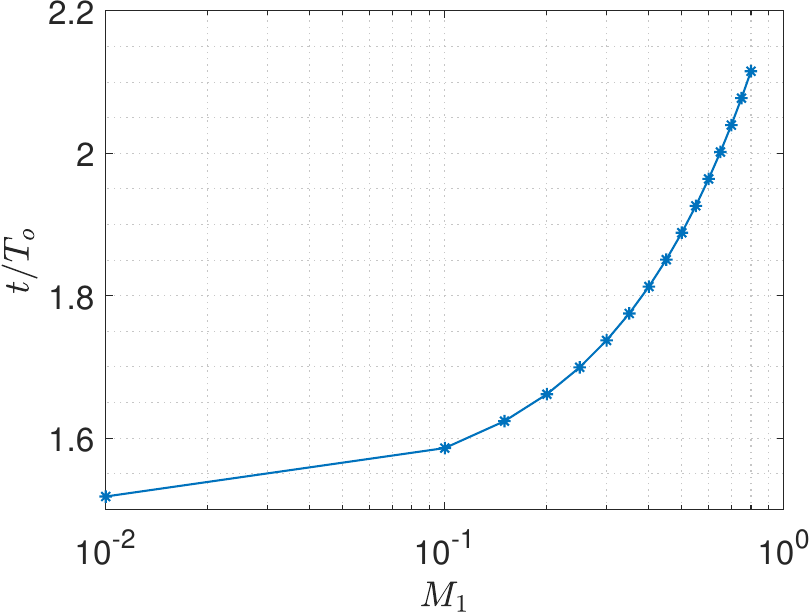}
\caption{Normal fluids (NFs)}
\end{subfigure}
\begin{subfigure}{0.45\linewidth}
\includegraphics[trim = 0 0 0 0,clip, width =65mm]{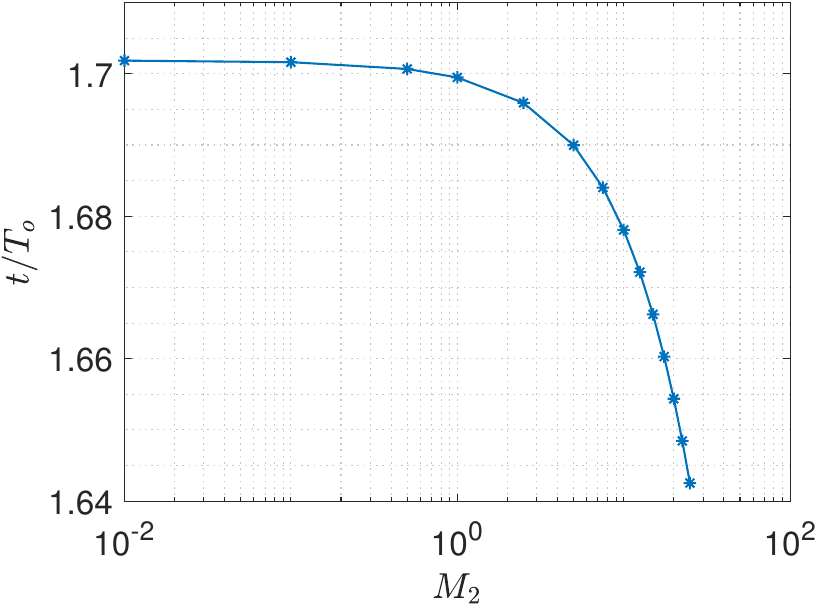}
\caption{Self-rewetting fluids (SRFs)}
\end{subfigure}
\caption{Dimensionless time duration $t/T_o$ for the onset of pinch-off for (a) normal fluids (NFs) with $M_2 = 0$ as a function of $M_1$ and (b) self-rewetting fluids (SRFs) with $M_1 = 0$ as a function of $M_2$. Here, $\mbox{Bo} = 0.1$, $\mbox{Oh} = 0.01$, $\tilde{\rho} = 1000$, and $\tilde{\mu} = 100$. Here, the time scale $T_o$ is given by $T_o = R/U_o = R/\sqrt{\sigma_o/(\rho_a R)}$.}
\label{NF_SRF_pinch_off_time}
\end{figure}

Finally, we study how changing $M_1$ and $M_2$ for NF and SRF, respectively, affects the time duration prior to the start of pinch-off when a nonuniform surface temperature is applied. To illustrate this, we plot in Fig.~\ref{NF_SRF_pinch_off_time} the duration for the onset of pinch-off for NFs and SRFs. Here, $\mbox{Bo} = 0.1$, $\mbox{Oh} = 0.01$, $\tilde{\rho} = 1000$, and $\tilde{\mu} = 100$ are the remaining parameters that we use in this study. It is seen that the time to pinch-off is significantly dependent on the choice of fluid.
With NFs,  an increase in the linear dimensionless surface tension sensitivity coefficient $M_1$ increases with an increase in the pinch-off time. This behavior contrasts with SRF, where the pinch-off time decreases with increasing the quadratic sensitivity coefficient $M_2$, which is consistent with the findings discussed earlier for a different boundary condition (see Fig.~\ref{NF_SRF_pinch_off_time_Q_01_and_02}). In summary, SRFs are more likely than NFs to pinch-off for a wide variety of conditions, while also speeding up the process.

\section{Summary and conclusions} \label{Sec.9 HF}

A temperature-dependent characteristic, surface tension is a property that is often important to the phenomenon of interfacial transport in fluids. Self-rewetting fluids (SRFs) have an unusual non-linear (quadratic) dependence on temperature, in contrast to normal fluids (NFs), which have a linear relationship between surface tension and temperature. A quadratic dependency with a minimum point and a positive gradient characterizes this relationship. As a result, SRFs have unique characteristics, including fluid movement towards surfaces at higher temperatures. Due to these special qualities, SRFs show great promise for various applications, including microfluidics, in terrestrial and microgravity environments.

The main goal of this study is to simulate and study the processes of a self-rewetting drop coalescing or pinching-off when it interacts with a self-rewetting liquid layer that is subject to nonuniform heating, which is achieved via imposing a Gaussian profile for the heat flux variation or nonuniform temperature distribution on its bottom surface. The central moment LB schemes for interface tracking, two-fluid motions, and energy transport were initially tested against an established benchmark problem involving coalescence/pinch-off during drop interface interactions for the non-heated case where prior experimental and numerical results are available to confirm its validity. We then investigated the pinch-off or coalescence mechanism of a drop on a liquid pool involving SRFs and comparing and contrasting their behavior with those of NFs. By studying the effects of the Bond number $\mbox{Bo}$ and the Ohnesorge number $\mbox{Oh}$ of SRF drops impinging on heated fluid layers, it is found that for small $\mbox{Oh}$, the pinch-off occurs because the surface tension forces dominates over the viscous forces. On the other hand, when we increase $\mbox{Oh}$, the pinch-off is effectively suppressed due to the greater influence of the counteracting viscous forces.

Furthermore, there is a striking distinction between the pinch-off process in NFs and SRFs; first, unlike NFs, where \emph{the pinch-off happens later}, SRFs experience \emph{the pinch-off process sooner}. Second, in the case of the SRFs, the fluids on the interface move toward the hotter zone, which is opposite to that in NFs. The Marangoni stress results from a positive (negative) surface tension gradient on the interface for SRFs (NFs), and this causes variations in thermocapillary flow fields between NFs and SRFs that explain the observed differences.

We then studied the effect of changing the linear and quadratic surface tension sensitivity coefficients $M_1$ and $M_2$, respectively compared to the non-heated case. We found that increase in $M_1$ decrease the critical value of $\mbox{Oh}$ at lower $\mbox{Bo}$ where the range of conditions for pinch-off to occur decreases in NFs. The opposite is observed in the SRF case where at higher quadratic coefficient $M_2$, the critical value of $\mbox{Oh}$ increases for lower $\mbox{Bo}$, on the pinch-off region in the $\mbox{Oh}$-$\mbox{Bo}$ map,increases, compared to the lower $M_2$ case because the resulting Marangoni forces promote the pinch-off process. Also, we investigated the effect of varying either $M_1$ or $M_2$ on the initiation of pinch-off in NFs and SRFs, respectively. The choice of fluid type significantly influences when the drop initiates the pinch-off process. In the case of NFs, the pinch-off time increases as the dimensionless linear surface tension coefficient $M_1$ increases, while the behavior is the opposite in SRFs, where the time to pinch-off decreases as the quadratic coefficient $M_2$ increases. These observations are maintained even when the heat flux boundary condition is replaced with an imposed nonuniform temperature condition on the bottom side of the liquid layer. Moreover, an increase in the imposed dimensionless heat flux $Q$ increases the Marangoni force with a concomitant reduction in time duration needed to pinch-off time in SRFs. In general, overall, it is found that in SRFs pinch-off across a wider range of parameters while also taking shorter time durations in this regard, which contrasts with the behavior observed in NFs. These findings may be utilized to devise strategies to manipulate coalescence/pinch-off processes by exploiting the unique thermocapillary phenomena that arise in the special class of SRFs under nonuniform heating. Furthermore, the results of this paper involved the assumption of the axial symmetry based on computational efficiency considerations, which, while realistic in various physical situations, can be relaxed to perform full three-dimensional simulations to represent the physical processes in more general settings.

\section*{Acknowledgements}
The authors would like to acknowledge the support of the US National Science Foundation (NSF) for research under Grant CBET-1705630. The third author (KNP) would also like to thank the NSF for support of the development of a computer cluster `Alderaan' hosted at the Center for Computational Mathematics at the University of Colorado Denver under Grant OAC-2019089 (Project ``CC* Compute: Accelerating Science and Education by Campus and Grid Computing''), which was used in performing the simulations.

\appendix

\section{Mapping Relations for the Central Moment LB Scheme on a D2Q9 lattice} \label{App B}
Here, we summarize the various mapping relations that are needed prior to and following the collision step, where different central moments are relaxed to their equilibria, in the central moment LB scheme on the D2Q9 lattice.

The transformation matrix $\tensr{P}$ mapping a vector of distribution functions $\mathbf{f}$ to a vector of raw moments $\bm{\kappa^{'}}$ is given by
\begin{equation}\label{eq:tensorP}
\tensr{P} = \begin{bmatrix}
     1  &\quad    1  &\quad    1  &\quad      1  &\quad    1  &\quad     1 &\quad    1  &\quad     1  &\quad     1 \\[10pt]
     0  &\quad    1  &\quad     0  &\quad    \um1  &\quad     0  &\quad    1 &\quad   \um1 &\quad   \um1  &\quad     1 \\[10pt]
     0  &\quad    0  &\quad     1  &\quad     0  &\quad   \um1  &\quad    1  &\quad     1  &\quad    \um1  &\quad  \um1 \\[10pt]
     0  &\quad    1  &\quad     0  &\quad     1  &\quad     0  &\quad    1  &\quad     1  &\quad     1  &\quad     1 \\[10pt]
     0  &\quad    0  &\quad     1  &\quad     0  &\quad     1  &\quad    1  & \quad    1  &\quad     1  &\quad     1 \\[10pt]
     0  &\quad    0  &\quad     0  &\quad     0  &\quad     0  &\quad    1  &\quad    \um1  &\quad     1  &\quad   \um1 \\[10pt]
     0  &\quad    0  &\quad     0  &\quad     0  &\quad     0  &\quad    1  &\quad     1  &\quad  \um1  &\quad  \um1 \\[10pt]
     0  &\quad    0  &\quad     0  &\quad     0  &\quad     0  &\quad    1  &\quad    \um1  &\quad  \um1  &\quad    1 \\[10pt]
     0  &\quad    0  &\quad     0  & \quad    0  &\quad     0  & \quad   1  &\quad     1  &\quad     1  &\quad    1
\end{bmatrix}
\end{equation}
Next, the transformation matrix $\tensr{F}$ mapping a vector of raw moments $\bm{\kappa^{'}}$ to a vector of central moments $\bm{\kappa}$ reads as
\begin{equation}\label{eq:tensorF}
\tensr{F}=
\begin{bmatrix}
      1  &    0  &    0  &     0  &    0  &     0 &    0  &     0  &     0 \\[10pt]

     \um u_x  &   1  &    0  &     0  &    0  &     0 &    0  &     0  &     0 \\[10pt]

      \um u_y  &    0  &   1  &     0  &    0  &     0 &    0  &     0  &     0 \\[10pt]

      u_x^2  &   \um 2u_x  &    0  &     1  &    0  &     0 &    0  &     0  &     0 \\[10pt]

      u_y^2  &    0  &    \um 2u_y  &     0  &    1  &     0 &    0  &     0  &     0 \\[10pt]

     u_x u_y  &   \um u_y  &    \um u_x  &     0  &    0  &     1 &    0  &     0  &     0 \\[10pt]

      \um u_x^2 u_y  &   2u_x u_y  &   u_x^2  &     \um u_y  &    0  &     \um 2u_x &    1  &     0  &     0 \\[10pt]

      \um u_x u_y^2 &    u_y^2  &    2u_x u_y  &     0  &    \um u_x  &    \um 2u_y &    0  &     1  &     0 \\[10pt]

      u_x^2 u_y^2  &    \um u_x u_y^2  &    \um u_x^2 u_y  &    u_y^2  &   u_x^2  &    4u_x u_y &    \um 2 u_y  &     \um 2  u_x  &     1 \\
\end{bmatrix}
\end{equation}
Then, the transformation matrix $\tensr{F}^{-1}$ mapping a vector of (post-collision) central moments $\bm{\tilde{\kappa}}$ to a vector of (post-collision) raw moments $\bm{\tilde{\kappa}}^{'}$ can be written as
\begin{equation}\label{eq:tensorFinverse}
\tensr{F}^{-1}=
\begin{bmatrix}
      1  &    0  &    0  &     0  &    0  &     0 &    0  &     0  &     0 \\[10pt]

      u_x  &   1  &    0  &     0  &    0  &     0 &    0  &     0  &     0 \\[10pt]

      u_y  &    0  &   1  &     0  &    0  &     0 &    0  &     0  &     0 \\[10pt]

      u_x^2  &   2u_x  &    0  &     1  &    0  &     0 &    0  &     0  &     0 \\[10pt]

      u_y^2  &    0  &    2u_y  &     0  &    1  &     0 &    0  &     0  &     0 \\[10pt]

     u_x u_y  &   u_y  &    u_x  &     0  &    0  &     1 &    0  &     0  &     0 \\[10pt]

      u_x^2 u_y  &   2u_x u_y  &   u_x^2  &     u_y  &    0  &     2u_x &    1  &     0  &     0 \\[10pt]

      u_x u_y^2 &    u_y^2  &    2u_x u_y  &     0  &    u_x  &    2u_y &    0  &     1  &     0 \\[10pt]

      u_x^2 u_y^2  &    u_x u_y^2  &    u_x^2 u_y  &    u_y^2  &   u_x^2  &    4u_x u_y &    2 u_y  &     2  u_x  &     1 \\
\end{bmatrix}
\end{equation}
It may be noted that if $\tensr{F}=\tensr{F}(u_x,u_y)$, then $\tensr{F}^{-1}=\tensr{F}(-u_x,-u_y)$ (see~\cite{yahia2021central}).

Finally, we express the transformation matrix $\tensr{P}^{-1}$ mapping a vector of (post-collision) raw moments $\bm{\tilde{\kappa}^{'}}$ to a vector of
(post-collision) distribution functions $\mathbf{\tilde{f}}$ as
\begin{equation}\label{eq:tensorPinverse}
\tensr{P}^{-1} =
\begin{bmatrix}
     1  &\quad    0  &\quad    0  &\quad    \um 1  &\quad   \um 1  &\quad     0 &\quad    0  &\quad     0  &\quad     1 \\[10pt]
     0  &\quad    \frac{1}{2}  &\quad     0  &\quad    \frac{1}{2}  &\quad     0  &\quad    0 &\quad   0 &\quad   \um \frac{1}{2}  &\quad     \um \frac{1}{2} \\[10pt]
     0  &\quad    0  &\quad     \frac{1}{2}  &\quad     0  &\quad   \frac{1}{2}  &\quad    0  &\quad     \um \frac{1}{2}  &    0  &\quad   \um \frac{1}{2} \\[10pt]
     0  &\quad    \um \frac{1}{2}  &\quad     0  &\quad     \frac{1}{2}  &\quad     0  &\quad   0 &\quad     0  &\quad     \frac{1}{2}  &\quad    \um \frac{1}{2} \\[10pt]
     0  &\quad    0  &\quad     \um \frac{1}{2}  &\quad     0  &\quad     \frac{1}{2}  &\quad    0  &\quad     \frac{1}{2} &     0  &\quad     \um \frac{1}{2} \\[10pt]
     0  &\quad    0  &\quad     0  &\quad     0  &\quad     0  &\quad    \frac{1}{4}  &\quad    \frac{1}{4}  &\quad     \frac{1}{4}  &\quad   \frac{1}{4} \\[10pt]
     0  &\quad    0  &\quad     0  &\quad     0  &\quad     0  &\quad    \um \frac{1}{4}  &\quad     \frac{1}{4}  &\quad  \um \frac{1}{4}  &\quad \frac{1}{4} \\[10pt]
     0  &\quad    0  &\quad     0  &\quad     0  &\quad     0  &\quad    \frac{1}{4}  &\quad    \um \frac{1}{4} &\quad  \um \frac{1}{4}  &\quad    \frac{1}{4} \\[10pt]
     0  &\quad    0  &\quad     0  &\quad     0  &\quad     0  &\quad    \um \frac{1}{4}  &\quad     \um \frac{1}{4}  &\quad     \frac{1}{4}  &\quad    \frac{1}{4}
\end{bmatrix}
\end{equation}


\end{document}